\documentclass[english]{svjour3}
\usepackage[T1]{fontenc}
\usepackage[latin9]{inputenc}
\usepackage{url}
\usepackage{amsmath}
\usepackage{amssymb}
\usepackage{graphicx}
\usepackage{physics}

\usepackage{ulem}

\usepackage{mathtools} 
\usepackage{hyperref}

\newcommand\nn{\nonumber}
\newcommand\la{\langle}
\newcommand\ra{\rangle}
\newcommand\f{\frac}
\newcommand\p{\partial}

\usepackage{color}
\definecolor{lightgray}{gray}{.90}%%%COLOR FOR GRAYBOX
\definecolor{darkred}{rgb}{0.9,0.1,0.1}%%%COLOR FOR SIDE COMMENT

\makeatletter
\RequirePackage{fix-cm}

\smartqed  % flush right qed marks, e.g. at end of proof
\usepackage{mathtools}
\makeatother
\graphicspath{{./Figures/}{./FiguresFinal/}}

\usepackage{babel}

\begin{document}
%\title{Quantum dynamics under continuous projective measurements: non-Hermitian description and the continuum space limit}
\title{
%Non-Hermitian description of continuous projective measurements on a quantum system ~~~~~~~OR~~
Quantum dynamics under continuous projective measurements: non-Hermitian description and the continuum space limit}
\titlerunning{Non-Hermitian description of continuous projective measurements}
\author{Varun Dubey \and C\'edric Bernardin \and Abhishek Dhar}
\authorrunning{Dubey, Bernardin and Dhar}

\institute{Varun Dubey \at International Centre for Theoretical Sciences, Tata Institute of Fundamental Research, Bengaluru 560089, India. 
\email{varun.dubey@icts.res.in} \and C\'edric Bernardin \at Universit\'e C\^ote d'Azur, CNRS, LJAD Parc Valrose, 06108 NICE Cedex 02, France. \email{Cedric.BERNARDIN@unice.fr}  \and Abhishek Dhar \at International Centre for Theoretical Sciences, Tata Institute of Fundamental Research, Bengaluru 560089, India. \email{abhishek.dhar@icts.res.in}}
\date{\today}
\maketitle
\begin{abstract}
The problem of the time of arrival of a quantum system in a specified state 
is considered in the framework of the repeated measurement protocol and in particular the limit of continuous measurements is discussed. It is shown that for a particular choice of system-detector coupling, the Zeno effect is avoided and the system can be described effectively by a non-Hermitian effective Hamiltonian. As a specific example we consider  the evolution of a quantum particle on a one-dimensional lattice that is subjected to position measurements at a specific site.  By solving the corresponding non-Hermitian wavefunction evolution equation, we present analytic closed-form results on the survival probability and the first arrival time distribution. Finally we discuss the limit of vanishing lattice spacing and show that this leads to a continuum description where the particle evolves via the free Schrodinger equation with complex Robin boundary conditions at the detector site. Several interesting physical results for this dynamics are presented.  

  %\PACS{PACS code1 \and PACS code2 \and more} \subclass{MSC code1 \and MSC code2 \and more}
\end{abstract}

\section{Introduction\label{sec:Introduction}}
The question of the time of arrival of a quantum particle at a detector is a difficult problem since this requires one to make measurements on a system which for  quantum systems  lead to a change in the system's evolution dynamics. The most dramatic manifestation of this fact is the so-called Zeno effect where a system under constant observation is never detected~\cite{misra1977}. In a more general setting, consider a quantum system that is prepared in an initial state $|\phi \ra$ and then evolves under a unitary dynamics. One is interested in the time of return of the system to the initial state or it's arrival  to a state $|\chi\ra$ that is orthogonal to $|\phi\ra$.  A well-defined protocol to answer this question is to consider a sequence of projective measurements  done on the system at regular intervals of time, say $\tau$,  whose output is $1$ (a ``yes'') if the system is in the desired final state and is otherwise  $0$ (a null measurement). The experiment stops when we record a $1$ and, if this happens on the $n$-th measurement, we say that the time of detection is $t=n \tau$. Till the time of detection the system's evolution thus consists of a unitary evolution that is conditioned by null measurements (non-detection in the target state) at regular intervals of time and hence repeated projections into the space complimentary to the target space. 

In the second of a series of three papers on the quantum time of arrival problem~\cite{allcock1969time}, Allcock discussed the repeated projection protocol in the context of the time of arrival of a particle which is released at some point $x<0$ on the real line and eventually detected by a measuring device at the origin $x=0$. Allcock considered a sequence of measurements at discrete time intervals which  project the particle to the domain $x<0$ till the time of detection. It was argued that this process could be effectively modeled by including an imaginary potential $-i V_0 \theta (-x)$, where $\theta(x)$ is the Heaviside function, in the free Schrodinger evolution of the particle. The strength of the potential, $V_0$, is related to the intervals between measurements, $\tau$, as $V_0 \sim \tau^{-1}$. The limit $\tau \to 0$ gives the Zeno effect. 
We are not aware of a  formal derivation of the equivalence of the repeated projection dynamics with the non-Hermitian Hamiltonian~\cite{echanobe2008,Halliwell2009,Halliwell2010}. In more recent work~\cite{dhar2015a,dhar2015b,lahiri2019} the first detection problem using the repeated measurement scheme has been investigated in a lattice model for a quantum particle  satisying the discrete Schrodinger equation. It was numerically demonstrated that a description of the repeated measurement dynamics by means of an effective non-Hermitian Hamiltonian is in fact quite accurate for small values of $\tau$ (more precisely for $\tau << \gamma^{-1}$ where $\gamma$ is the transition rate between lattice sites). In an independent set of work~\cite{bach2004one,grunbaum2013}, for systems with discrete Hilbert spaces, the arrival time problem using the repeated measurements protocol has been treated using a renewal approach similar to what one uses for the classical Polya first passage problem. Using this approach, a number of interesting physical results on first passage in the   lattice Schrodinger problem have been obtained in Refs.~\cite{friedman2016,friedman2017,thiel2018,thiel2018spectral,thiel2020uncertainty}. In particular it has been shown that the return probability on an infinite one dimensional lattice is less than one and we do not have the Polya recurrence of the 1-D random walk. The distribution of first detection times has been shown to scale as $1/t^3$. Different aspects of the equivalence to the non-Hermitian Hamiltonian have been explored in Refs.~\cite{lahiri2019,thiel2020,thiel2020quantization}. Other interesting results include discussions of the non-Hermitian description in many-body systems~\cite{elliott2016quantum,kozlowski2016non} and in the context of experiments~\cite{barontini2013controlling,ashida2016quantum}.

The present work makes several contributions. First we show that, for systems with a discrete Hilbert space, the equivalence  between the repeated projection dynamics and the non-Hermitian Hamiltonian can be established rigorously in a particular limit where we let the measurement time interval $\tau \to 0$ while taking a large value (of order $1/\tau^{1/2}$) for the strength of the coupling strength between the system and detector. This choice is motivated by similar ones used in the description of the trajectories of continuously monitored quantum systems~\cite{BauerBenoistBernard2012}. Secondly we study the example of a quantum  particle on different $1D$ lattices (finite, semi-infinite and infinite) and, using a Fourier-Laplace transform of the effective equations of motion, obtain closed-form expressions for the survival probability and detection time distribution. We compute the spectrum of the effective non-Hermitian Hamiltonian and show that the results from the Fourier-Laplace solution also follow from a  spectral analysis.  Finally, we consider the lattice-to-continuum space limit and show that the evolution is described by a Schrodinger equation  with complex Robin boundary condition at the location of the detector.  Apart from the survival probability and the full spectrum,  the form of the ``surviving'' wave-function is also discussed.
 Such boundary conditions have earlier been discussed, somewhat formally,  in the context of the time of arrival problem~\cite{werner1987,TumulkaTeufel2019,Tumulka2019}.

The plan of the paper is as follows. In Sec.~\eqref{sec:mapping} we describe the limiting procedure which gives  the exact mapping between the repeated projective measurements and the non-Hermitian Hamiltonian. In Sec.~\eqref{sec:Examples} we consider the example of a quantum particle on  three $1D$ lattices and write the form of the effective non-Hermitian Hamiltonian for these cases. The  exact results for survival probability and detection time distribution using the Fourier-Laplace transform approach and the Green's function approach are presented in Sec.~\eqref{sec:surv}. In Sec.~\eqref{sec:continuum} we discuss the continuum space limit and finally conclude with a summary of the results in Sec.~\eqref{sec:conclusions}.

\section{Description of dynamics under repeated projective measurements by
an effective non-Hermitian Hamiltonian: exact mapping}
\label{sec:mapping}
Consider a quantum system whose states belong to the Hilbert space $\mathcal{H}$. We assume that $\mathcal H = \mathcal S\oplus \mathcal D$ can be written as the sum of two orthogonal complementary subspaces, where $\mathcal S$ is a ``system'' space and ${\mathcal D}$ a ``detector'' space. Let  $\left|i\right\rangle $ be an orthonormal basis of $\mathcal S$ and $\left|\alpha\right\rangle $ an orthonormal basis of $\mathcal D$. The set of states $\{\left|i\right\rangle ,\left|\alpha\right\rangle \}$ together form a complete orthonormal basis of $\mathcal H$. Hence we have
\begin{equation}
\label{eq:orthonormality0}
\sum_{i}\left|i\right\rangle \left\langle i\right| + \sum_{\alpha}\left|\alpha\right\rangle \left\langle \alpha\right| = I
\end{equation}
where $I$ is the identity operator. The most general Hamiltonian describing such a system is given
by
\begin{align}
H&=H^{(\mathcal S)}+H^{(\mathcal D)}+H^{(\mathcal{SD})},  \\
H^{(\mathcal S)}&=\underbrace{\sum_{i,j}H_{ij}^{(S)}\left|i\right\rangle \left\langle j\right|}_{\text{System}},~~~~ 
H^{(\mathcal D)}=\underbrace{\sum_{\alpha,\beta}H_{\alpha\beta}^{(D)}\left|\alpha\right\rangle \left\langle \beta\right|}_{\text{Detector}},  \\ 
 H^{(\mathcal{SD})}&=\underbrace{\sum_{i,\alpha}\left[H_{i\alpha}^{(\mathcal{SD})}\left|i\right\rangle \left\langle \alpha\right|+H_{\alpha i}^{(\mathcal{SD})}\left|\alpha\right\rangle \left\langle i\right|\right]}_{\text{System-Detector}}. 
\end{align}
Let $\left|\psi(t)\right\rangle \in {\mathcal H} $ denote the state of the total system
at time $t$. The unitary evolution of this state with the above Hamiltonian $H$
is given by
\begin{equation*}
\left|\psi(t)\right\rangle =U_{t}\left|\psi(0)\right\rangle ,\,\,\,\,U_{t}=\exp\left(-\imath tH/\hbar \right).
\end{equation*}
Consider instantaneous projective measurements made to find if the system is
in the detector subspace $\mathcal{D}$. This corresponds to the projection
operator
\begin{equation}
P=\sum_{\alpha}\left|\alpha\right\rangle \left\langle \alpha\right|
\end{equation}
while the complementary operator corresponding to projection into
the system subspace $\mathcal{S}$ is given by
\begin{equation*}
Q=I-P=\sum_{i}\left|i\right\rangle \left\langle i\right|.
\end{equation*}
Now imagine an experiment where initially the system starts in the
space $\mathcal{S}$ and begins to evolve unitarily. We keep making
measurements, at regular intervals of time $\tau$, to find if the system
has arrived in the detector subspace $\mathcal{D}$. If the result of the measurement is negative, the system (now projected back into $\mathcal{S}$) continues its unitary evolution, until the next measurement
and the process is repeated. The experiment stops when we get a positive
result indicating arrival into $\mathcal{D}$. Thus the evolution
of the state vector consists of a sequence of steps in $\mathcal S$, each consisting
of a unitary evolution followed by the projection $Q$. Looking at
the state of the system conditioned on survival (non-detection), let
us denote by $\left|\psi(n\tau)\right\rangle $ the state just after the
$n$-th measurement. Then it can be shown~\cite{dhar2015b} that
\begin{equation}
\left|\psi(n\tau)\right\rangle =\widetilde{U}_{\tau}^{n} \left\vert \psi(0)\right\rangle, \quad \widetilde{U}_{\tau}=Q U_{\tau} Q.
\end{equation}
The probability, $S(n\tau)$, that the system stays undetected after
the $n$-th measurement is given by the norm of the state
\begin{align}
S(n\tau)=\left\langle \psi(n\tau) \vert \psi(n\tau)\right\rangle . \label{SP}
\end{align}
The normalized state after the $n$-th measurement is given by
\begin{equation}
\left|\widetilde{\psi}(n\tau)\right\rangle =\frac{\left|\psi(n\tau)\right\rangle }{\sqrt{S(n\tau)}}.
 \label{normpsi}
\end{equation}

It was numerically demonstrated  in Refs.~\cite{dhar2015a,dhar2015b,lahiri2019} that for the case where the time between measurements $\tau$ is small (compared to typical time scales in the unitary evolution) the above dynamics is accurately described by a continuous time evolution with a non-Hermitian effective Hamiltonian.  However, to obtain a description with continuous time, we need the limit $\tau \to 0$ and this is problematic because it leads to the Zeno effect.  
Here we discuss a systematic approach where this problem is overcome and the effective non-Hermitian description 
can be obtained in a rigorous way. For this we let the system-detector couplings scale with the detection interval $\tau$
as
\begin{equation*}
H_{i\alpha}^{({\mathcal{SD}})}=\sqrt{\frac{\gamma_{i\alpha}}{\tau}},\,\,\,H_{\alpha i}^{({\mathcal{SD}})}=\sqrt{\frac{\gamma_{\alpha i}}{\tau}}.
\end{equation*}

Further, denoting $H_{ij}^{(\mathcal S)}=\gamma_{ij}$ and $H_{\alpha\beta}^{(\mathcal D)}=\gamma{}_{\alpha\beta}$
we note the form of the Hamiltonian $H$

\begin{equation}
H=\underbrace{\sum_{i,j}\gamma_{ij}\left|i\right\rangle \left\langle j\right|}_{H^{\mathcal S}}+\underbrace{\sum_{\alpha,\beta}\gamma{}_{\alpha\beta}\left|\alpha\right\rangle \left\langle \beta\right|}_{H^{\mathcal D}}+\underbrace{\sum_{i,\alpha}\left[\sqrt{\frac{\gamma_{i\alpha}}{\tau}}\left|i\right\rangle \left\langle \alpha\right|+\sqrt{\frac{\gamma_{\alpha i}}{\tau}}\left|\alpha\right\rangle \left\langle i\right|\right]}_{H^{\mathcal{SD}}}.
\label{eq:Princ-Hamil}
\end{equation}
We assume from now on that $H$ is measured in units of $\hbar$ and so has units of frequency, then the $\gamma$ coefficients (assumed
to be positive) so do too. Expanding $\widetilde{U}_\tau$ in a power series,
one has

\[
\widetilde{U}_\tau=Q\exp\left(-\imath\tau H\right)Q=QQ+\left(-\imath\tau\right)QHQ+\frac{\left(-\imath\tau\right)^{2}}{2}QHHQ+\dots
\]
Note that $\left|\psi(n \tau)\right\rangle $ belongs to $\mathcal{S}$,
which means that $Q\left|\psi(n\tau)\right\rangle =\left|\psi(n\tau)\right\rangle $. The following identities follow
from the orthonomality of the basis states Eq.~(\ref{eq:orthonormality0}) and the definition of $Q$:
\begin{align}
QH^{(\mathcal S)}Q& =H^{(\mathcal S)},\,\,QH^{(\mathcal D)}Q=0,\,\,QH^{(\mathcal{SD})}Q=0,\\
Q\left[H^{(\mathcal S)}\right]^{2}Q &=\left[H^{(\mathcal S)}\right]^{2},\,\,Q\left[H^{(\mathcal D)}\right]^{2}Q=0,\\
Q\left[H^{(\mathcal{SD})}\right]^{2}Q&=\sum_{i,j,\alpha}H_{i\alpha}^{({\mathcal {SD}})}H_{\alpha j}^{({\mathcal {SD}})}\left|i\right\rangle \left\langle j\right|=\frac{1}{\tau}\sum_{i,j,\alpha}\left|i\right\rangle \left\langle j\right|\sqrt{\gamma_{i\alpha}\gamma_{\alpha j}}.
\end{align}
By use of these in the expansion of $\widetilde{U}_\tau$, we then get
\begin{equation}
\widetilde{U}_\tau=I-\imath\tau H^{\mathcal S}-\tau V^{\mathcal S}+\mathcal{O}(\tau^{2})\label{eq:Uexp}
\end{equation}
where
\begin{equation}
V_{ij}^{\mathcal S}=\frac{1}{2}\sum_{\alpha}\sqrt{\gamma_{i\alpha}\gamma_{\alpha j}} 
\label{Veff}
\end{equation}

\medskip

We now take the continuum limit $\tau\to0$, $n\to\infty$ while keeping
$t=n\tau$ finite. Let $\left|\psi(t)\right\rangle =\left|\psi(n\tau)\right\rangle $.
Then, using Eq.~(\ref{eq:Uexp}), we write
\[
\imath\frac{\partial\left|\psi(t)\right\rangle }{\partial t}=\lim_{\tau\to0}\imath\frac{\left|\psi(n\tau+\tau)\right\rangle -\left|\psi(n\tau)\right\rangle }{\tau}=\imath\frac{\widetilde{U}\left|\psi(n\tau)\right\rangle -\left|\psi(n\tau)\right\rangle }{\tau}
\]
and we get a new Shrodinger's equation with an effective non-Hermitian Hamiltonian $H^{\text{eff}}$
\begin{equation}
\imath\frac{\partial\left|\psi(t)\right\rangle }{\partial t}=H^{{\rm eff}}\left|\psi(t)\right\rangle , \quad H^{\text{eff}}=H^{\mathcal S}-\imath V^{\mathcal S}.
\label{eq:EffEvol}
\end{equation}
Thus we have shown that the repeated-measurement dynamics, in the
special limit where the time interval between measurements $\tau$ goes to
zero while the system-detector interaction diverges like $\tau^{-1/2}$, is given by the
non-Hermitian Hamiltonian in Eq.~(\ref{eq:EffEvol}). Note that the
dynamics is completely evolving in the system subspace $\mathcal{S}$. \\

Note that in the limit $\tau \to 0$ the survival probability at time $t$ is obtained from Eq.~\eqref{SP}: 
\begin{equation}
\label{eq:surv-prob-cont}
S(t)= \left\langle \psi (t) | \psi (t) \right\rangle,
\end{equation}
with $\vert \psi (t) \rangle$ solution of the effective Shrodinger's equation Eq.~(\ref{eq:EffEvol}). The distribution of first detection times is given by
\begin{align}
F(t)=\f{d S}{d t}=-2 \la \psi|V^{\mathcal S}|\psi \ra. \label{eq:fpt}
\end{align}

\section{Example of a quantum particle on different one dimensional lattice}
\label{sec:Examples}

%\begin{figure}
%\centering
%\includegraphics[scale=0.9]{Figures/lattice-crop.pdf}
%\caption{\label{fig:1-d Lattice}Finite 1-d lattice with Detector D at site0}
%\end{figure}

We now consider some simple physical examples  of the Hamiltonian Eq.~(\ref{eq:Princ-Hamil}) corresponding to the case of  a single particle  hopping on a one-dimensional lattice $\Lambda \subset \mathbb Z$ containing the site $0$ and subjected to the action of a potential $v(n)$. %In Fig.~\eqref{fig:1-d Lattice} we show schematically a  quantum particle moving on a finite lattice. 
A detector performs a position measurement at the site $0$ after every time interval of length $\tau$. The basis vectors $|i \ra$ now correspond to the position eigenstates and we have $P=|0\ra \la 0 |$. We consider three different choices of lattices which we specify now along with the corresponding Hamiltonian:
\begin{enumerate}
\item Finite lattice $\Lambda=\{0,\ldots, N\}$ of size $N\ge 2$:
\begin{equation}
\begin{split}
H=&- \gamma_0 \sum_{n=2}^{N}\left(~\left|n\right\rangle \left\langle n-1\right|+\left|n-1\right\rangle \left\langle n\right| -2 \left| n\right\rangle \left\langle n \right|~ \right) + (2+\beta) \gamma_0 \left|1\right\rangle \left\langle 1\right|\\
&-\sqrt{\f{\gamma}{\tau}} \left(~\left|0\right\rangle \left\langle 1\right|+ \left|1\right\rangle \left\langle 0\right|~\right).
\end{split}
\label{eq:effh-finite1}
\end{equation}
\item Semi-infinite lattice  $\Lambda={\mathbb N}$
\begin{equation}
\begin{split}
H=&- \gamma_0 \sum_{n=2}^{\infty}\left(~\left|n\right\rangle \left\langle n-1\right|+\left|n-1\right\rangle \left\langle n\right| -2 \left| n\right\rangle \left\langle n \right|~\right) + (2+\beta) \gamma_0 \left|1\right\rangle \left\langle 1\right|\\
&-\sqrt{\f{\gamma}{\tau}}  \left(~\left|0\right\rangle \left\langle 1\right|+ \left|1\right\rangle \left\langle 0\right|~\right).
\label{eq:effh-N2}
\end{split}
\end{equation}
\item Infinite lattice $\Lambda=\mathbb Z$: 
\begin{equation}
\begin{split}
H=
&-\sum_{n\in\mathbb{Z}\setminus\{0\}}\left[~(1-\delta_{n,1})\left|n\right\rangle \left\langle n-1\right|+(1-\delta_{n,-1})\left|n\right\rangle \left\langle n+1\right| -2 \left| n\right\rangle \left\langle n \right|~\right] \\
&+ (2+\beta) \gamma_0 \left[~ \left|1\right\rangle \left\langle 1\right| + \left|-1\right\rangle \left\langle -1\right|~ \right]\\
&-\sqrt{\f{\gamma}{\tau}} \left[ ~\left|0\right\rangle \left\langle 1\right|+ \left|1\right\rangle \left\langle 0\right|+
\left|0\right\rangle \left\langle -1\right|+ \left|-1 \right\rangle \left\langle 0\right| ~\right].
\label{eq:effh-Z3}
\end{split}
\end{equation}
\end{enumerate}
We introduce the dimensionless parameter $\alpha=\frac{\gamma}{2\gamma_{0}}$ which can be regarded as the strength of the measurement. We have assumed  that the potential $v(n)$ is constant and equal to $2 \gamma_0$ except on the two sites $-1, +1$ where it is equal to $(\beta+2)\gamma_0$. Hence $\beta$ is also a dimensionless parameter measuring the impurity of the potential near the detector. These parameters are capsulated in the complex number $w=\alpha+\imath \beta.$
Denote then by 
\begin{equation*}
H_{\Lambda}=\cfrac{H^{\text{eff}}}{\gamma_0}
\end{equation*}
the corresponding scaled effective Hamiltonian. By using Eqs.~(\ref{Veff},\ref{eq:EffEvol}) we get the following effective Hamiltonians corresponding to the system-detector Hamiltonians in Eqs.~(\ref{eq:effh-finite1},\ref{eq:effh-N2},\ref{eq:effh-Z3}):
\begin{enumerate}
\item Finite lattice $\Lambda_N=\{1,\ldots, N\}$ of size $N\ge 2$:
\begin{equation}
H_{\Lambda_N}=- \sum_{n=2}^{N}\big[\left|n\right\rangle \left\langle n-1\right|+\left|n-1\right\rangle \left\langle n\right| -2 \left| n\right\rangle \left\langle n \right| \big]+ (2-\imath w )\left|1\right\rangle \left\langle 1\right|.
\label{eq:effh-finite}
\end{equation}
\item Semi-infinite lattice $\mathbb N$:
\begin{equation}
H_{\mathbb N}=-\sum_{n=2}^{\infty}\big[\left|n\right\rangle \left\langle n-1\right|+\left|n-1\right\rangle \left\langle n\right| -2 \left| n\right\rangle \left\langle n \right|\big] +(2-\imath w)  \left|1\right\rangle \left\langle 1\right|.
\label{eq:effh-N}
\end{equation}
\item Infinite lattice $\mathbb Z$: 
\begin{equation}
\begin{split}
H_{\mathbb{Z}}=
&-\sum_{n\in\mathbb{Z}\setminus\{0\}}\big[(1-\delta_{n,1})\left|n\right\rangle \left\langle n-1\right|+(1-\delta_{n,-1})\left|n\right\rangle \left\langle n+1\right| -2 \left| n\right\rangle \left\langle n \right|\big] \\
&-\imath w\sum_{n,n'\in\{-1,1\}}\vert n\rangle\,  \langle n' \vert.
\label{eq:effh-Z}
\end{split}
\end{equation}
\end{enumerate}

Observe that these operators are not Hermitian (nor normal) and therefore a priori not diagonalisable. In fact for e.g. $N=2$ we can explicitly show that for specific values of $w$ the operator $H_{\Lambda_2}$ is not diagonalisable. We discuss the spectral properties of the Hamiltonian $H_{\mathbb N}$ and $H_{\mathbb Z}$ in App.~\eqref{sec:app-spec-g}.

\section{Calculation of Survival Probability}
\label{sec:surv}
Having mapped the measurement problem to that of non-unitary evolution given by Eq.~\eqref{eq:EffEvol},
the methods developed in Ref.~\cite{krapivsky2014} can be employed to calculate
the survival probability. The three cases considered in the previous section are discussed in the following sub-sections.

\subsection{The lattice $\Lambda_{N}$}

The Hamiltonian $H_{\Lambda_N}$ can be rewritten as 
\begin{equation*}
H_{\Lambda_N}= -\Delta_N -\imath w \vert 1\rangle\, \langle 1\vert
\end{equation*}
where $\Delta_N$ is the discrete Laplacian on $\Lambda_N$. We first prove that in this case the survival probabilty goes to $0$ with an exponential decay at large times. We recall that the spectrum of the Laplacian ($\epsilon_q=-2(1-\cos (q) )$, where $q=s \pi/(N+1),~s=1,2,\ldots,N$. Hence it is contained in $\{z \in {\mathbb C}\; ; \; {\Re}(z) <0, \; \Im(z)=0\}$. Let $\lambda$ be an eigenvalue of $-\imath H_{\Lambda_N}$ associated to some eigenvector $|\psi\rangle$ then we have that
\begin{equation*}
{\lambda} \langle \psi \vert \psi \rangle = \imath  \, \langle \psi \vert \Delta_N \vert \psi\rangle -{w} \, \vert \langle 1 \vert \psi\rangle \vert^2 .
\end{equation*}
If $\langle 1 \vert \psi \rangle \ne 0$, $\Re ( \lambda)<0$ because $\Re (w) >0$. If $\langle 1 \vert \psi \rangle= 0$ then $\imath \Delta_N \vert \psi \rangle =\lambda \vert \psi\rangle$, hence $\vert \psi \rangle$ is an eigenvector of $\Delta_N$ associated to the eigenvalue $-\imath \lambda$. Since the (explicit) eigenvectors of $\Delta_N$ are not such that $\langle 1 \vert \psi \rangle= 0$, the latter case never holds. We conclude that the spectrum of $-\imath H_{\Lambda_N}$ is contained in  $\{z \in {\mathbb C}\; ; \; {\Re}(z) <0\}$. By using the Jordan-Chevalley decomposition of $-\imath H_{\Lambda_N}$ we conclude that for any $|\psi (0)\rangle$, 
\begin{equation*}
S(t)=\langle \psi (t) \vert \psi(t) \rangle = O(e^{-\mu t})
\end{equation*}
for some real $\mu>0$. Hence the survival probability $S(t)$ goes to $0$ as $t\to \infty$ exponentially fast. We will see in the two next subsections that when the lattice is infinite, it is no longer the case.  For the case where $-\imath H_{\Lambda_N}$ has a non-degenerate spectrum, let $\lambda_m$ be the eigenvalue with the smallest negative real part. Then the decay constant is given by $\mu=-\Re [\lambda_m]$. The ``surviving'' wave function at large times would be given by $|\psi(t)\ra \sim e^{\lambda_m t} |\chi_m \ra$ where $ |\chi_m \ra$ is the eigenvector of $-\imath H_{\Lambda_N}$ corresponding to the eigenvalue $\lambda_m$.

From Eq.~\eqref{eq:fpt} we see that the first passage time distribution is given by $F(t)=-\tfrac{\partial S}{\partial t} (t)=2 \alpha |\psi_1(t)|^2$, where we use the notation $\psi_i=\la i |\psi\ra$. 
We now show that one can write a formal solution for $\psi_1(t)$, by using the information on the eigenspectrum of the Hermitian part of the effective Hamiltonian, which in this case is the lattice Laplacian. We first write the equation of motion:
\begin{align}
\imath \f{\p |\psi\ra}{\p t} = -\Delta_N |\psi\ra -i w \la 1|\psi\ra |1\ra.
\end{align}
Taking a Laplace transform $|\tilde{\psi}(s)\ra= \int_0^\infty dt e^{- s t} |\psi(t)\ra$, we get
\begin{align}
-|\psi(0)\ra +s |\tilde{\psi}(s)\ra = \imath \Delta_N |\tilde{\psi}(s) \ra - w \la 1|\tilde{\psi} (s) \ra |1\ra.
\end{align}
Defining the Green's function $G(s)=[s-i \Delta_N]^{-1}$, we get the following formal solution
\begin{align}
|\tilde{\psi}(s)\ra = G(s) |\psi(0)\ra -  w \la 1|\tilde{\psi} (s) \ra G(s) |1\ra.
\end{align}
Assuming an initial wavefunction, $|\psi(0)\ra = |\ell \ra$, localized at a site $\ell$ we take a projection of the above equation on the state $|1\ra$ to get the Laplace transform ${\mathcal L} \psi_1$ of $\psi_1$:
\begin{align}
[\mathcal{L}\psi_1](s):=\la 1|\tilde{\psi}(s)\ra = \f{G_{1 \ell}(s)}{1+ w  G_{11}(s)}, \label{eq:LTpsi1}
\end{align}
where the matrix elements $G_{ij}(s)=\la i | G(s)|j \ra$ can be written in terms of the eigenfunctions $\phi_q(j)=\sqrt{2/(N+1)} \sin (q j)$ of the Laplacain $\Delta_N$, and corresponding eigenvalues $\epsilon_q = -2(1-\cos q)$ with $q=s\pi/(N+1)$, $s=1,2,\ldots,N$. One gets
\begin{align}
G_{ij} (s)= \sum_q \f{\phi_q(i) \phi_j(q)}{s-i \epsilon_q}
\end{align}
and hence an explicit expression for the Laplace transform $({\mathcal L} \psi_1 )(s)$. By using an inverse Laplace transform formula we can then get an explicit formula for the first passage time distribution $F(t)$ which remains however difficult to exploit, even qualitatively. Similar expressions have recently been discussed in Ref.~\cite{thiel2020} who also point the analogy with the renewal approach for the repeated measurement problem.  In the next section we will make use of the analogous formula as  Eq.~\eqref{eq:LTpsi1} to get explicit results on first passage distribution on the infinite lattice. For the finite lattice we  now present some numerical results on the form of the survival probability.\\  

\begin{figure}
\includegraphics[scale=.55]{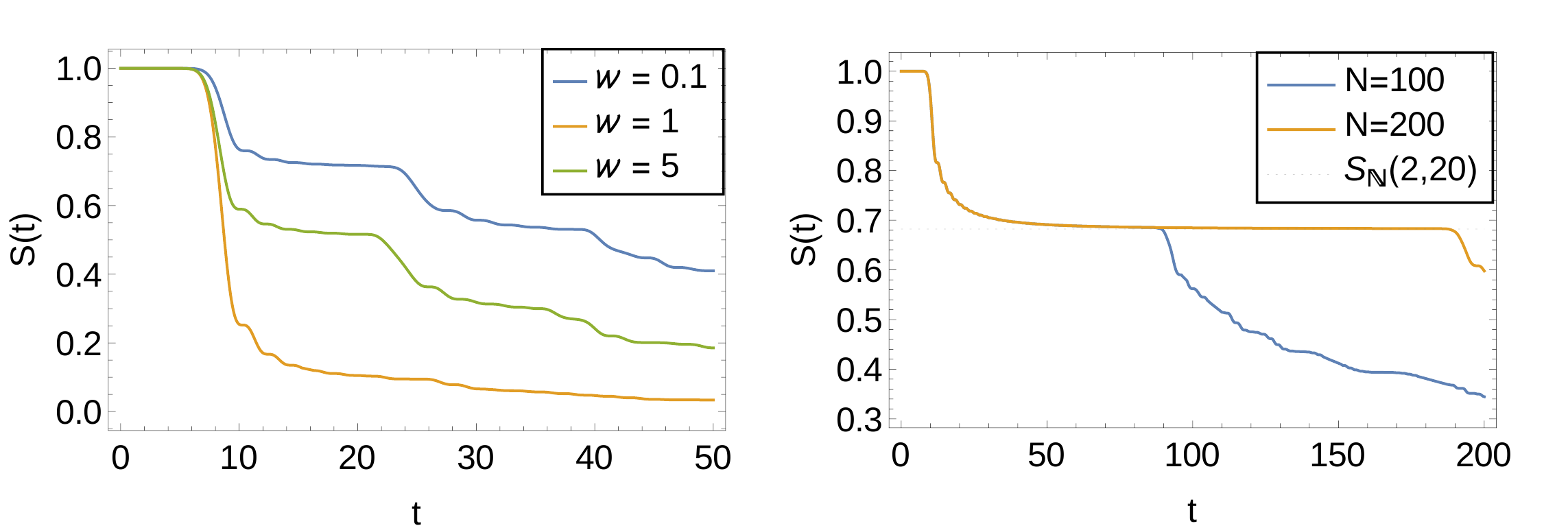}
\caption{\label{fig:1-d Simul} In the graph on the left, the survival probability $S(t)$ is plotted for $N=15$, $\psi_{i}(0)=\delta_{i,15}$ and for different values of $w$. In the graph to the right, the survival probability $S(t)$ is plotted for lattice sizes $N=100$ and $N=200$. In both cases $\psi_{i}(0)=\delta_{i,20},\,w=2$. The dashed line is the value of survival probability $S(\infty)$ obtained from Eq.~\eqref{eq:S_N} for the $\mathbb{N}$ Lattice.}
\end{figure}

{\bf Numerical results:} The numerical results are obtained by a direct solution of the non-Hermitian Schrodinger equation. As shown above in this case we have $\left[S(t)\right]_{t\rightarrow\infty}=0$ and we always eventually  detect the particle.
In  Fig.~\eqref{fig:1-d Simul}  (left panel) we plot the
decay of survival probability on a system for which $N=15$ for three different values of $w$.  The survival
probability cascades to $0$ over time. We observe a non-monotonic dependence with $S(t)$ decaying slowly for both  very small $w=0.1$ as well large $w=5$.   The several plateaues in the survival probability can be understood as arising from ballistic propagation of the particle with a 
a group velocity $\approx 2$, such that as the wave packet hits the detector, the surival probability plunges considerably. The  velocity $2$ corresponds to the maximum group velocity $d\epsilon_q/dq$. 
In  Fig.~\eqref{fig:1-d Simul} (right panel) we show the decay of survival probability
for relatively larger systems. In all cases, $\psi_{i}(0)=\delta_{i,20}$
and $w=2.0$. In this case the first short plateau in $S(t)$ corresponds to the time taken for the wave-packet to reach the detector while the second longer plateau corresponds to the wave-packet traveling from the origin to the right end and back. Hence this second time-scale can be seen to be proprtional to the lattice size $N$. The dashed line indicates the infinite time survival probability $S_{\mathbb{N}}(w,20)$ on the semi-infinite lattice ${\mathbb N}$ which is computed exactly  in the next section (see Eq. \eqref{eq:S_N}). Similar finite size effects have been discussed recently in Ref.~\cite{lahiri2019}.

\subsection{$\mathbb{N}$ Lattice}

The Schrodinger equation corresponding to the effective Hamiltonian $H_{\mathbb{N}}$ given by Eq.~(\ref{eq:effh-N}) is
\begin{equation}
\imath\cfrac{\partial \psi_{n}}{\partial t}=
\begin{dcases}
(2-\imath w)\psi_{1}-\psi_{2}, & n=1,\\
2 \psi_n -\psi_{n-1}-\psi_{n+1},& n\ge 2.
\label{eq:SE-NLattice}
\end{dcases}
\end{equation}
By introducing the fictitious Dirichlet boundary condition $\psi_0 (t)=0$ the Eq.~\eqref{eq:SE-NLattice} can be written as
\begin{equation}
\label{eq:SE-NLattice-bis}
\imath\cfrac{\partial \psi_{n}}{\partial t}= 2 \psi_n -\psi_{n-1}-\psi_{n+1} -\imath w \delta_1 (n) \psi_1 ,\quad  n\ge 1.
\end{equation}
Decompose the wave function on the following orthonormal sin-basis
\begin{gather*}
\psi_{n}(t)=\sqrt{\dfrac{2}{\pi}}\int_{0}^{\pi}dk\,\hat{\psi}(k,t)\sin(n\,k) \quad \text{with} \quad \hat{\psi}(k,t)=\sqrt{\dfrac{2}{\pi}}\sum_{n=1}^{\infty}\psi_{n}(t)\,\sin(n\,k). 
\end{gather*}
Observe that the Dirichlet boundary condition is automatically satisfied. Assume $\psi_{n}(0)=\delta_{n,n_{0}}$ so that $\hat{\psi}(k,0)=\sqrt{\dfrac{2}{\pi}}\sin\left(n_{0}k\right)$. Then it follows from Eq.~\eqref{eq:SE-NLattice-bis} that 
\begin{equation}
\imath\cfrac{\partial \hat{\psi}}{\partial t} (k,t)-2(1-\cos\left(k\right)) \hat{\psi}(k,t)=- \imath w \,  \sqrt{\dfrac{2}{\pi}}\,  \psi_{1}(t) \, \sin\left(k\right).
\label{eq:Nlattice-eq-bef}
\end{equation}
The Laplace transforms of $\hat{\psi}(k,t)$ and $\psi_{1}(t)$ are by definition given by
\begin{gather}
\tilde{\psi}(k,s)=[\mathcal{L} \hat{\psi}(k,\cdot)] (s)=\int_{0}^{\infty}dt\,\exp(-st)\hat{\psi}(k,t),\\
[\mathcal{L} \psi_{1}] (s)=\int_{0}^{\infty}dt\,\exp(-st)\,\psi_{1}(t)=\sqrt{\dfrac{2}{\pi}}\int_{0}^{\pi}dk\,\tilde{\psi}(k,s)\,\sin(k).
\label{eq:linkpsi1-Psi}
\end{gather}
Taking the Laplace Transform of Eq.~(\ref{eq:Nlattice-eq-bef})
and by use of the above equations
\begin{equation}
\tilde{\psi}(k,s)=\imath\sqrt{\dfrac{2}{\pi}}\, \frac{1}{\imath s-2 \left(1-\cos\left(k\right) \right)}\left\{\sin\left(n_{0}k\right)-w  \sin\left(k\right) \left[\mathcal{L} \psi_{1}\right](s)\right\}.
\label{eq:linkpsi1-Psi2}
\end{equation}
The two relations Eq.~(\ref{eq:linkpsi1-Psi}) and Eq.~(\ref{eq:linkpsi1-Psi2}) between $\tilde{\psi}(k,s)$ and $[\mathcal{L}\psi_{1}](s)$
above give
\begin{equation}
\left[\mathcal{L}\psi_{1}\right] (s)=\imath \dfrac{\displaystyle \dfrac{2}{\pi}\int_{0}^{\pi}dk\,\dfrac{\sin\left(k\right)\sin\left(n_{0}k\right)}{\imath s-2(1-\cos (k)) }}{1+\imath\dfrac{2w }{\pi}\displaystyle\int_{0}^{\pi}dk\,\dfrac{\sin^{2}k}{\imath s-2(1-\cos (k))}}.
\label{eq:LapPsi1}
\end{equation}
Note that this also follows from Eq.~\eqref{eq:LTpsi1} on taking the limit $N \to \infty$. 
The integrals involved in Eq.~\eqref{eq:LapPsi1} can be evaluated
by means of contour integration for any complex number $s$ such that ${\Re} (s)>0$, see App.~\eqref{subsubsec:integral}. We get then that
\begin{equation}
\left[\mathcal{L}\psi_{1}\right] (s)=
=-\imath\frac{\imath^{n_0}\bigg[-\left(\frac{s}{2}+\imath\right)+\sqrt{\left(\frac{s}{2}+\imath\right)^{2}+1}\bigg]^{n_0}}{1+ w \bigg[-\left(\frac{s}{2}+\imath\right)+\sqrt{\left(\frac{s}{2}+\imath\right)^{2}+1}\bigg]}
\label{eq:s-spacePsi1}
\end{equation}
where  $\sqrt z$ denotes the principal square root of $z\in \mathbb C\backslash {\mathbb R}^*_{-}$ (using the nonpositive real axis as a branch cut). We have hence obtained an explicit expression of the Fourier-Laplace transform $\tilde{\psi}$ of the wave function $\psi$ by plugging Eq.~\eqref{eq:s-spacePsi1} into Eq.~\eqref{eq:linkpsi1-Psi2}.

\medskip 
We now turn to the computation of the survival probability. Recalling Eq.~(\ref{eq:surv-prob-cont}), the survival probability after time $t$ is $S(t)=\sum_{n=1}^{\infty}\left|\psi_{n}(t)\right|^{2}$. From Eq.~\eqref{eq:SE-NLattice}, one has
\begin{equation*}
\imath\dfrac{\partial}{\partial t}\left|\psi_{n}\right|^{2}=
\begin{dcases}
-2 \imath \Re (w) \left|\psi_{1}\right|^{2}-\psi_{1}^*\psi_{2}+\psi_{1}\psi_{2}^*, & n=1,\\
-\psi_{n}^*\psi_{n-1}+\psi_{n}\psi_{n-1}^* - \psi_{n}^*\psi_{n+1}+\psi_{n}\psi_{n+1}^*, & n\ge 2.
\end{dcases}
\end{equation*}
Since $\psi_{n}$ goes to $0$ as $n\rightarrow\infty$, the
equations can be summed to obtain ${d S}/{d t}= -2\Re (w)\left|\psi_{1}\right|^{2}.$
Integrating, we get the survival probability 
\begin{equation}
S_{\infty}=\lim_{t\to \infty} S(t) =1-2\Re(w) \int_{0}^{\infty}dt\,\left|\psi_{1}(t)\right|^{2}.
\label{eq:SurvInteg}
\end{equation}
For square integrable function $\psi_1$, we have ~\cite{Schiff}, Chapter 4, that
\begin{equation}
\int_{0}^{\infty}dt\,\left|\psi_{1}(t)\right|^{2}=\lim_{\epsilon \to 0^+} \dfrac{1}{2\pi\imath}\int_{\epsilon-\imath\infty}^{\epsilon+\imath\infty}ds\,\left|[\mathcal{L} \psi_{1}](s) \right|^{2}.
\label{eq:int-Laplace-tr}
\end{equation}
For $s=\epsilon+2 \imath (x-1)$ in the limit $\epsilon\rightarrow0^{+}$, by Eq.~\eqref{eq:s-spacePsi1}, we have that
\begin{equation}
\label{eq:squareLapTr}
\left|[\mathcal{L} \psi_{1}](s) \right|^{2}=
\begin{cases}
\dfrac{\left(x+\sqrt{x^{2}-1}\right)^{2n_0}}{1+\left|w\right|^{2}\left(x+\sqrt{x^{2}-1}\right)^{2}+2\Im(w)\left(x+\sqrt{x^{2}-1}\right)},  \quad x < -1, \\
\\
\dfrac{1}{1+\left|w\right|^{2}+2\big[\Re(w)\sqrt{1-x^{2}}+\Im(w)\,x\big]}, \quad -1 \le x  \le 1, \\
\\
\dfrac{\left(x-\sqrt{x^{2}-1}\right)^{2n_0}}{1+\left|w\right|^{2}\left(x-\sqrt{x^{2}-1}\right)^{2}+2\Im(w)\left(x-\sqrt{x^{2}-1}\right)}, \quad x>1.
\end{cases}
\end{equation}\\

Denote by  $S_{\mathbb N}(w, n_{0})$ the survival probability given by Eq.~(\ref{eq:SurvInteg}) corresponding to the initial condition $\vert \psi (0) \rangle =\delta_{n_{0}}$, when the strength measurement parameter is $\alpha={\Re} (w)$ and the impurity parameter is $\beta={\Im} (w)$. From Eq.~(\ref{eq:SurvInteg}) and Eqs.~(\ref{eq:int-Laplace-tr}, \ref{eq:squareLapTr}), after simplification, we obtain then that
\begin{equation}
\begin{split}
S_{\mathbb N}(w,n_{0})= 1&-\frac{{\Re}(w)}{\pi\left|w\right|}\int_{-\frac{\pi}{2}}^{\frac{\pi}{2}}\frac{\cos\,\theta}{\frac{1}{2}\left(\left|w \right|+\frac{1}{\left|w\right|}\right)+\cos\left(\theta-\varphi\right)}d\theta\\
&-\frac{2{\Re}(w)}{\pi\,}\int_{0}^{1}\frac{u^{2n_0-2}(1-u^{2})(1+\left|w\right|^{2}u^{2})}{(1+\left|w\right|^{2}u^{2})^{2}-(2{\Im}(w)\,u)^{2}}du
\end{split}
\label{eq:S_N}
\end{equation}
where $\varphi={\rm{Arg}} (w)$. Only the second integral is depending on $n_0$. Both can be explicitly computed but the formula obtained do not have a simple form. For large $n_0$ a simple expansion can be done for the latter at any order, see App.\eqref{subsec:Integrale approximation}. Therefore for $n_{0}\gg1$, we have e.g. that
\begin{equation}
\label{eq:saddleappforSP}
\begin{split}
S_{\mathbb N}(w,n_{0})\approx 1&-\frac{{\Re}(w)}{\pi\left|w\right|}\int_{-\frac{\pi}{2}}^{\frac{\pi}{2}}\frac{\cos\,\theta}{\frac{1}{2}\left(\left|w\right|+\frac{1}{\left|w\right|}\right)+\cos\left(\theta-\varphi\right)}d\theta\\
&-\cfrac{1}{n_0^2} \;\frac{{\Re}(w)}{\pi} \dfrac{ (1+| w|^2)}{(1+|w|^2)^2 -4 [{\Im} (w)]^2}.
\end{split}
\end{equation}

\begin{figure}
	\includegraphics[scale=0.55]{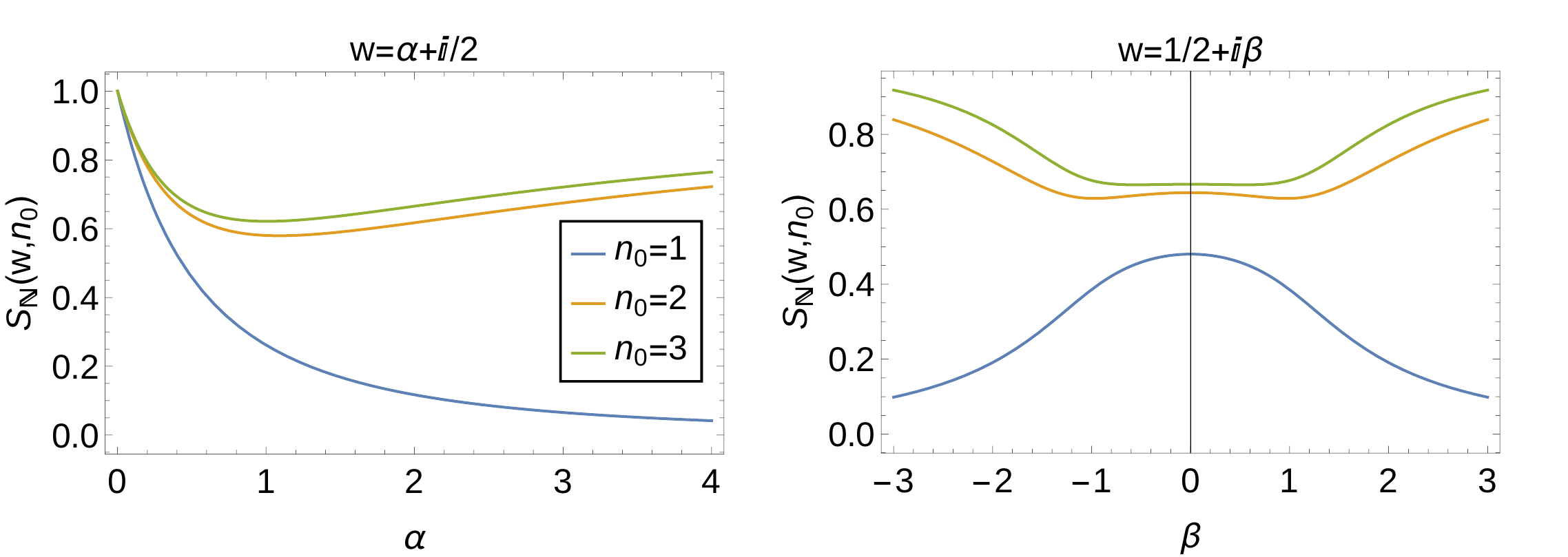}
	\caption{\label{fig:SvsAlpha}Plots showing variation of $S_{\mathbb N}$ (Eq.~\eqref{eq:S_N}) with $w$ for various starting positions $n_{0}$. The left plot is for $w=\alpha + \imath/2$ and the right plot for $w=1/2+\imath\beta$ }
\end{figure}

In fact we can also get an explicit formula for the function $\psi_1$ and therefore for the survival probability $S(t)=1-2\Re(w) \int_{0}^{t} dt'\,\left|\psi_{1}(t')\right|^{2}$ at time $t$. For that it is sufficient to recognize in Eq.~\eqref{eq:s-spacePsi1} the Laplace transform of some explicit function. In App.~\eqref{subsec:psi_1-expression} we identify this function by using properties of the Bessel functions of the first kind: 
\begin{equation}
\label{eq:Bessel}
J_{k}(t)=\cfrac{1}{2\pi}\int_{-\pi}^\pi dx\, e^{-\imath(kx -t\sin(x))}.
\end{equation}
We find  that 
\begin{equation}
\label{eq:psi_1Bessel} 
\psi_1 (t) = -2 \imath^{n_0+1} e^{-2\imath t} \sum_{k=0}^{\infty} (k+n_0) (- w)^{k}\dfrac{J_{k+n_0} (2t)}{2t},
\end{equation}
which immediately gives us the explicit form of the first detection distribution $F(t)= 2 \Re [w] |\psi_1 (t)|^2$.  At large times using the asymptotic form of $J_k(2 t)$ we get
\begin{equation}
\Rightarrow\psi_{1}(t)\asymp\frac{(-\imath)^{n_{0}+1}e^{-\imath2t}}{2\sqrt{\pi t^{3}}}\Bigg[e^{\imath\left(2t-\frac{\pi}{4}+\frac{\imath n_{0}\pi}{2}\right)}f(n_{0},w)+e^{-\imath\left(2t-\frac{\pi}{4}+\frac{\imath n_{0}\pi}{2}\right)}\left[f(n_{0},w^{*})\right]^{*}\Bigg]\label{eq:psi1asymptotic}
\end{equation}
where 
\[
f(n_{0},w)=\frac{n_{0}-\imath(n_{0}-1)w}{(1-\imath w)^{2}}.
\]
Hence we get the long time decay form $F(t) \sim 1/t^3$.

%\begin{figure}
%\includegraphics[scale=0.6]{Figures/SvsAlpha}
%\caption{\label{fig:SvsAlpha}Survival Probability for $n_{0}\in\{1,2,3\}$}
%\end{figure}
%
%Numerical evaluation from Equation (\ref{eq:S_N}) gives $S(2,20)=0.682431$
%which confirms the simulation results shown in Figure \ref{fig:1-d Larger Sites}.
%Figure \ref{fig:SvsAlpha} shows the variation of Survival probability
%$S (\alpha, n_0)$ with $\alpha$ for various starting positions $n_{0}$. For $n_{0}=1$,
%the survival probability goes to $0$ with increasing strength of
%measurement. For $n_{0}>1$ the survival probabilities attain minimum
%at around $\alpha\approx1$ as is evident from visual inspection.
%In fact the function $b\to\left[\frac{4}{\pi}\frac{b^{2}+1}{b^{2}-1}\arctan\frac{b-1}{b+1}\right]$
%has a minimum at $b=1$.

\subsection{$\mathbb{Z}$ Lattice}

The calculations here are along the same lines as for the $\mathbb{N}$
Lattice case. For the effective Hamiltonian $H_{\mathbb{Z}}$ given by Eq.~(\ref{eq:effh-Z}) the
Schrodinger equation is equivalent to 
\begin{equation}
\label{eq:shro_Z}
\begin{split}
&\imath\cfrac{\partial \psi_{n}}{\partial t}=2\psi_n -\psi_{n+1}-\psi_{n-1} -\imath w (\delta_{-1} (n)+\delta_{1} (n)) (\psi_{-1} +\psi_1), \quad n\in {\mathbb Z}\backslash\{0\},\\
& \psi_0 (t) =0.
\end{split}
\end{equation} 
Let us define 
\begin{gather*}
\hat{\psi}^\pm (k,t)=\sqrt{\dfrac{2}{\pi}}\sum_{n=1}^{\infty}\psi_{\pm n}(t)\,\sin(n\,k) \quad \Rightarrow \quad \psi_{\pm n}(t)=\sqrt{\dfrac{2}{\pi}}\int_{0}^{\pi}dk\,\hat{\psi}^{\pm} (k,t)\sin(n\,k)
\end{gather*}
which satisfy both the same equation
\begin{equation}
\label{eq:Zlattice-eq-bef}
\imath\cfrac{\partial \hat{\psi}^{\pm}}{\partial t} (k,t)-2(1-\cos\left(k\right)) \hat{\psi}^{\pm}(k,t)=- \imath w  \sqrt{\dfrac{2}{\pi}}\,  [\psi_{1}(t)+\psi_{-1} (t)]  \, \sin\left(k\right).
\end{equation}
We assume that $\psi_n(0) =\delta_{n,n_0}$ with $n_0\ge 1$ so that ${\hat \psi}^{+} (k, 0)=\sqrt{\dfrac{2}{\pi}}\sin\left(n_{0}k\right)$ and ${\hat \psi}^- (k,0)=0$. The Laplace transforms of $\hat{\psi}^{\pm}(k,t)$ and $\psi_{\pm 1}(t)$ are
by definition
\begin{gather}
\tilde{\psi}^{\pm} (k,s)=[\mathcal{L} \hat{\psi}^{\pm}(k,\cdot)] (s)=\int_{0}^{\infty}dt\,\exp(-st)\hat{\psi}^{\pm}(k,t),\\
[\mathcal{L} \psi_{\pm1}] (s)=\int_{0}^{\infty}dt\,\exp(-st)\,\psi_{\pm1}(t)=\sqrt{\dfrac{2}{\pi}}\int_{0}^{\pi}dk\,\tilde{\psi}^{\pm} (k,s)\,\sin(k).
\end{gather}
Taking the Laplace Transform of Eq.~\eqref{eq:Zlattice-eq-bef} and by use of the above equations
\begin{equation}
\begin{split}
\tilde{\psi}^{+}(k,s)&=\imath\sqrt{\dfrac{2}{\pi}}.\frac{1}{\imath s-2 \left(1-\cos\left(k\right) \right)}\left\{\sin\left(n_{0}k\right)-w  \sin\left(k\right) \mathcal{L} [\psi_{-1}+\psi_{1}](s)\right\},\\
\tilde{\psi}^{-}(k,s)&= - \imath \sqrt{\dfrac{2}{\pi}}.\frac{1}{\imath s-2 \left(1-\cos\left(k\right) \right)}  w \sin\left(k\right) \mathcal{L} [\psi_{-1}+\psi_{1}] (s).
\end{split}
\end{equation}
Then we get
\begin{equation}
\left[\mathcal{L}(\psi_{-1}+\psi_1) \right] (s)=\imath \, \dfrac{\displaystyle \dfrac{2}{\pi}\int_{0}^{\pi}dk\,\dfrac{\sin\left(k\right)\sin\left(n_{0}k\right)}{\imath s-2(1-\cos (k)) }}{1+\imath\dfrac{4w }{\pi}\displaystyle\int_{0}^{\pi}dk\,\dfrac{\sin^{2}k}{\imath s-2(1-\cos (k))}}
\label{eq:LapPsi11}
\end{equation}
from which follows the explicit expression for the Fourier-Laplace transform of the wave function. This is the same expression as Eq.~\eqref{eq:LapPsi1} where $w$ has been replaced by $2w$. \\

Proceeding as in derivation of Equation Eq.~(\ref{eq:SurvInteg}), in
this case the expression for the survival probability is 
\begin{equation}
S_{\infty}=\lim_{t\to \infty} S(t)= 1-2\Re (w) \int_{0}^{\infty}dt\,\left|\psi_{-1}(t)+\psi_{1}(t)\right|^{2}.\label{eq:SurIntegZ}
\end{equation}
If $S_{\mathbb Z} (w,n_{0})$ denotes the survival probability $S_\infty$ corresponding to this initial condition $\psi_{n}(0)=\delta_{n,n_{0}}$ we get that 
\begin{equation}
\begin{split}
S_{\mathbb Z} (w,n_{0})= S_{\mathbb N} (2w, n_0).
\end{split}
\label{eq:S_Z}
\end{equation}

%
%
%Numerical evaluation of Equation (\ref{eq:S_Z}) gives $S(0.1,5)=0.903799,\,S(0.5,5)=0.81313,\,$
%$S(2,5)=0.886983$. These values compare well with simulations in
%Figure \ref{fig:LargerZfigure}.

\subsection{Green's Function approach}
\label{sec:Green}

It is of interest to recover the value Eq.~(\ref{eq:S_N}) of the survival probability and the value Eq.~\eqref{eq:psi_1Bessel} of $\psi_1 (t)$ by considering the spectral analysis of $H_{\mathbb{N}}$ performed in App.~\eqref{subsec:sa-H_N}. A similar analysis can be performed on the lattice $\mathbb Z$  as in App.~\eqref{subsec:sa-H_Z}.

We recall that $w=\alpha+\imath \beta$ and we denote $\zeta=\tfrac{\beta - \imath \alpha}{1+\beta -\imath \alpha}= \tfrac{w}{\imath +w}$. It is proved in this appendix that the Hamiltonian $H_{\mathbb N}$ has a complete ``real orthonormal'' basis of eigenfunctions. This basis is composed of scattering states Eq.~\eqref{eq:etank} $\{\eta^k\; ; \; k\in (0,\pi)\}$ (associated with the eigenvalue $E_s(k)=2(1-\cos (k))$, and, if $|1-\xi^{-1}|<1 \Leftrightarrow |w|>1$, supplemented by a bound state $\eta^b$ Eq.~\eqref{eq:etanb}, associated to the eigenvalue $E_b (\zeta)=1+\zeta^{-1}$.\\

From this spectral analysis we have that the solution $\psi (t)$ of Eq.~\eqref{eq:SE-NLattice} with initial condition $\psi (0)$ can be expressed as
\begin{equation}
\psi (t) = G(t) \psi(0)
\end{equation}
where the entries of the infinite matrix $G(t)$ are given by
\begin{equation}
G_{n m} (t) = \int_{0}^\pi dk \, \eta_n^k \eta_m^k\,  e^{-\imath E_s (k) t} \; +\; {\bf 1}_{\vert w \vert >1} \, \eta_n^b\eta_m^b\,  e^{-\imath E_b (\zeta) t}.  
\end{equation}
In particular, starting with an initial condition $\vert \psi (0) \rangle =\delta_{n_0}$ we have that $\psi_1 (t) = G_{1, n_0} (t)$. By using the explicit forms  of $\{\eta^k\; ; \; k\in (0,\pi)\}$, $E_s (k)$ in Eq.~\eqref{eq:etank} and using that $\zeta/(1-\zeta) = -\imath w$ we get
\begin{equation*}
\eta_1^k \eta_{n_0}^k\,  e^{-\imath E_s (k) t}=-\cfrac{\imath}{\pi} e^{-2 \imath (1-\cos(k))t} \, \sin (k) \left[\cfrac{e^{\imath n_0 k}}{1-\imath w e^{\imath k}} - \cfrac{e^{-\imath n_0 k}}{1-\imath w e^{-\imath k}} \right].
\end{equation*}
Assuming $|w|<1$  we can transform the previous expression in series since $\vert\imath w e^{\pm \imath k}\vert<1$, exchange the sum with the integrals, and by using symmetries we get
\begin{equation}
\label{eq:tat-r}
\int_{0}^\pi dk \, \eta_n^k \eta_m^k\,  e^{-\imath E_s (k) t}= -\cfrac{\imath}{\pi} \sum_{p=0}^\infty (\imath w)^p \int_{-\pi}^\pi dk\, e^{-2 \imath (1-\cos(k))t} \, \sin (k) e^{\imath (n_0+p)k}.
\end{equation}
By performing an integration by parts in the definition of the Bessel function $J_p$ (see Eq.~\eqref{eq:Bessel}) we get
\begin{equation*}
\int_{-\pi}^\pi dk\, \sin(k) e^{2\imath t \cos (k)} e^{\imath p k} =\cfrac{ \pi \imath^{p}}{t} p J_{p} (2t)
\end{equation*}
and plugging this in Eq.~\eqref{eq:tat-r} we recover Eq.~\eqref{eq:psi_1Bessel}. 

If $|w|>1$ we have to take into account the existence of a bound state but for the rest we proceed in a similar way by rewriting for $|z|=1$
\begin{equation*}
\cfrac{1}{1-\imath w z}=\cfrac{\imath}{w z} \, \cfrac{1}{1+\tfrac{\imath}{w z}}= \cfrac{\imath}{w z} \,  \sum_{p=0}^\infty (-1)^p \, \left( \tfrac{\imath}{w z}\right)^p.
\end{equation*}
Then we get the same expression as before  Eq.~\eqref{eq:psi_1Bessel} for $\psi_1 (t)$. The expression for $|w|=1$ is obtained from the previous expression by a continuity extension.

\section{Continuum limit}
\label{sec:continuum}

In this section we discuss the continuum limit of the infinite models defined in Section \eqref{sec:Examples} when the lattice spacing $\epsilon$ goes to $0$.

\subsection{Half-line case}

We consider  the Hamiltonian $H_{\mathbb{N}}$ given by Eq.~\eqref{eq:effh-N}:
%\begin{align}
%H_{\mathbb{N}}^{m}= \gamma_0 \sum_{n=1}^{\infty}\left[\left|n\right\rangle \left\langle n+1\right|+\left|n+1\right\rangle \left\langle n\right| - 2 \left|n\right\rangle \left\langle n\right|  \right] +(v-\imath\gamma/2)  \left|1\right\rangle \left\langle 1\right|,~ \label{eq:effh-finite}
%\end{align}
\begin{equation*}
H_{\mathbb N}=-\sum_{n=1}^{\infty}\big[\left|n\right\rangle \left\langle n+1\right|+\left|n+1\right\rangle \left\langle n\right| -2 \left| n\right\rangle \left\langle n \right|\big] -\imath w \left|1\right\rangle \left\langle 1\right|.
\end{equation*}
We recall that Schrodinger equation take the form
\begin{equation*}
\imath\cfrac{\partial \psi_{n}}{\partial t}=
\begin{dcases}
(2-\imath w)\psi_{1}-\psi_{2}, & n=1,\\
2 \psi_n -\psi_{n-1}-\psi_{n+1},& n\ge 2.
\end{dcases}
\end{equation*}
These  are equivalent to the equations
\begin{equation}
\imath\cfrac{\partial \psi_{n}}{\partial t}=
2\psi_n-\psi_{n+1}-\psi_{n-1}, ~~ n \geq 1, \label{eq:SEmp}
\end{equation}
with the boundary condition $\psi_0= \imath w  \psi_{1}$ which can be rewritten in the form
\begin{equation}
\label{eq:zeta-psi0}
\psi_0 - \dfrac{\imath w}{1-\imath w}(\psi_1-\psi_0) =0.  
\end{equation}
We now define a small lattice spacing parameter $\epsilon$ and choose  $w$ such that  
\begin{align}
\label{eq:zeta00}
\frac{\imath w}{\imath w-1} = \frac{\zeta}{\epsilon}, 
\end{align}
where $\zeta$ is a finite complex number. We define the continuous wave function $\Psi$ as {\footnote{We assume that the initiale wave fonction $\psi_n (0)$ satisfies this property.}} 
$$\Psi(x , t) = \lim_{\epsilon\to 0} \epsilon^{-1/2} \psi_{[x/\epsilon]} \big(t \epsilon^{-2}\big).$$
Then in the limit $\epsilon\to 0$ Eqs.~(\ref{eq:SEmp},\ref{eq:zeta-psi0}) reduce  to   
\begin{equation}
\begin{gathered}
\imath \frac{\partial\Psi}{\partial t} =- \frac{\partial^2\Psi}{\partial x^2}, \quad \text{with complex Robin b.c.}\quad \left[\Psi+\zeta\tfrac{\partial\Psi}{\partial x} \right]_{x=0}=0.
\label{eq:SE-cont}
\end{gathered}
\end{equation}
Note that to obtain the desired limit in Eq.~\eqref{eq:zeta00} we need to set $\alpha =\Re(w)= \mathcal{O}(\epsilon)$ and $\beta=\Im{w}=-1+\mathcal{O}(\epsilon)$. Hence from the definition of $\zeta$ and the fact that $\alpha=\Re(w) >0$ we observe that  ${\Im} (\zeta)<0$ while $\Re (\zeta)$ can have either sign. Robin boundary conditions for real $\zeta$ have been studied and they arise in models of generic reflecting walls~\cite{allwright2016robin}. Here we have $\zeta$ as a complex number.

We now proceed with the solution of the above boundary value problem with a specified initial wave function $\Psi_0 (x) = \Psi(x , 0)$. The eigenspectrum of the system in Eq.~\eqref{eq:SE-cont} is discussed in App.~\eqref{subsec:app-spec-halfline}, where it is shown that for $\Re (\zeta) < 0$, the spectrum only has scattering eigenfunctions while for  $\Re (\zeta) > 0$, one has in addition a bound state that is localized near the origin. In this section we consider this latter case, the discussion would be similar for the other case.
The scattering functions $\{\eta^k\; ; \; k>0\}$ and the bound state $\eta^b$ defined respectively by Eq.~\eqref{eq:eta_k} and Eq.~\eqref{eq:eta_b} in App. \eqref{subsec:app-spec-halfline} form a complete basis of eigenfunctions of the Laplacian operator on the half line with the complex Robin boundary condition at the origin, see Eq.~\eqref{eq:SE-cont}. Moreover they satisfy the ``real orthogonality relations'' Eq.~\eqref{eq:orthonormality}. Then we can write the general time-dependent solution as
\begin{equation}
\begin{split}
\Psi(x , t)&= \int_0^\infty dk~ c(k) \eta^k(x) e^{- \imath k^2 t}  + c_b \eta^b(x) e^{\imath t/\zeta^2},~ \label{gensol} \\
{\rm where}\quad c(k)&= \int_0^\infty dx ~\Psi_0(x) \eta^k(x), \quad c_b= \int_0^\infty dx ~\Psi_0(x) \eta^b(x).
\end{split}
\end{equation}
The coefficients $c(k)$ as well as the oscillating integrals above are well defined for any initial wave function $\Psi_0$ which is sufficiently smooth since then the $c(k)$'s have a fast decay in $k$.  If $\Psi_0 \in {\mathbb L}^2 ([0,\infty))$ is only square integrable the previous formula has to be understood by using an approximation of the initial condition by smooth initial conditions. The approximation scheme propagates in time thanks to the decay Eq.~\eqref{eq:fptd} of the ${\mathbb L}^2$-norm. 

From the explicit formula Eq.~\eqref{gensol} and performing a standard saddle point approximation (see App.~\eqref{subsec:DecayFPTD} details) we  find the non-universal asymptotic form
\begin{equation}
\label{eq:Asymptotic}
\lim_{t\to \infty} {\sqrt {2t}}\,  \Psi (2ty , t) =  - c (y) \; \dfrac{\left(1-\imath\zeta y \right)}{\sqrt{1+\zeta^{2} y^2}} \exp\left[\imath\left( y^2 t-\tfrac{\pi}{4}\right)\right].
\end{equation}
Equivalently we can say that in the limit $t \to \infty$, for any finite  $x/t$, the wavefunction tends to the limiting form $\Psi \to \Psi_\infty$, where 
\begin{equation}
\Psi_\infty(x,t) = -\frac{1}{\sqrt{2t}} c\big({\tfrac{x}{2t}}\big) \; \dfrac{\left(1-\imath\zeta \tfrac{x}{2t} \right)}{\sqrt{1+\zeta^{2} \tfrac{x^2}{4t^2}}} \exp\left[\imath\left( \tfrac{x^2}{4t}-\tfrac{\pi}{4}\right)\right].  \label{eq:scaling}
\end{equation}
For $x\lesssim \sqrt{t}$ we need the form of $c(k)$ at $k \to 0$. Using the ecxplicit form of the wavefunctions in App.~\eqref{subsec:app-spec-halfline} we find immediately that 
\begin{align}
c(k) \sim \f{2 \imath k }{\sqrt{2 \pi}} m_{\Psi_0}, 
~~{\rm where} ~m_{\Psi_0}  =\int_0^{\infty} dx \, (x-\zeta) \Psi_0 (x)
\end{align}
Hence we have for $x \lesssim \sqrt{t}$
\begin{align}
\Psi_\infty(x,t) \approx m_{\Psi_0}  \frac{x}{\sqrt{4 \pi t^3}} \exp\left[\imath\left( \tfrac{x^2}{4t}+\tfrac{3\pi}{4}\right)\right], \label{eq:scal0}
\end{align}
which has a universal structure apart from the factor $m_{\Psi_0}$.

\begin{figure}
	\includegraphics[scale=0.37]{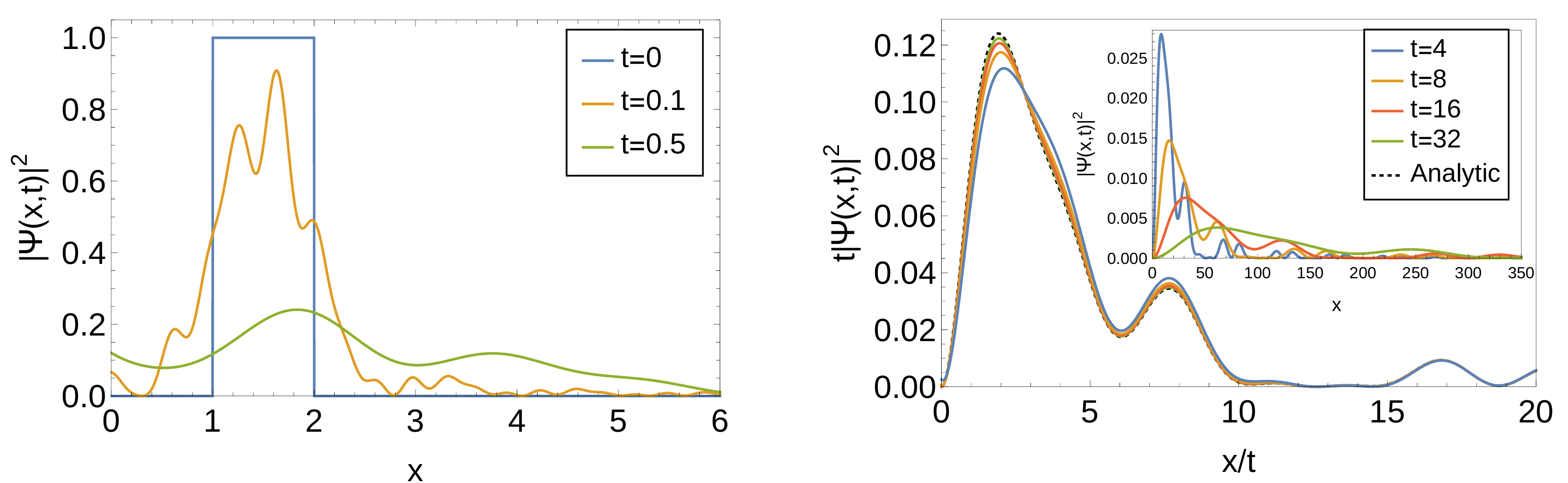}
	\caption{\label{fig:wavesq}
		Plot showing  $|\Psi(x,t)|^2$ at times $t=0,0.1,0.5$, obtained from the solution in Eq.~\eqref{gensol} with the square initial condition $\Psi(x,0)=1$ for $1<x<2$ and zero elsewhere. The parameter value $\zeta=0.2-0.5 i$ was taken.} 
\end{figure}

 As an explicit numerical example, we now consider the evolution of an initial wavefunction  of the form
\begin{align*}
\Psi_0(x)=1,~~{\rm for}~1<x<2, 
\end{align*}
and zero elsewhere. In this case the basis expansion coefficients are given: 
\begin{equation*}
\begin{split}
c(k)&=\frac{1}{\sqrt{2\pi(1 + k^2 \zeta^2)}} \frac{\sin (k/2) }{k/2}\bigg[k \zeta \cos (3 k/2) - \sin (3 k/2)\bigg],\\
c_b&=\sqrt{2 \zeta} e^{-2/\zeta} (-1+e^{1/\zeta}).
\end{split}
\end{equation*}
We choose $\zeta=0.2-0.5 \imath$. In Fig.~\eqref{fig:wavesq} we how the evolution of the wavefunction at early (a) and late times (inset of b). For the the scaled wave function in Fig.~\eqref{fig:wavesq}b and we see an excellent agreement with the analytic form in Eq.~\eqref{eq:Asymptotic}.

\begin{figure}
	\includegraphics[scale=.6]{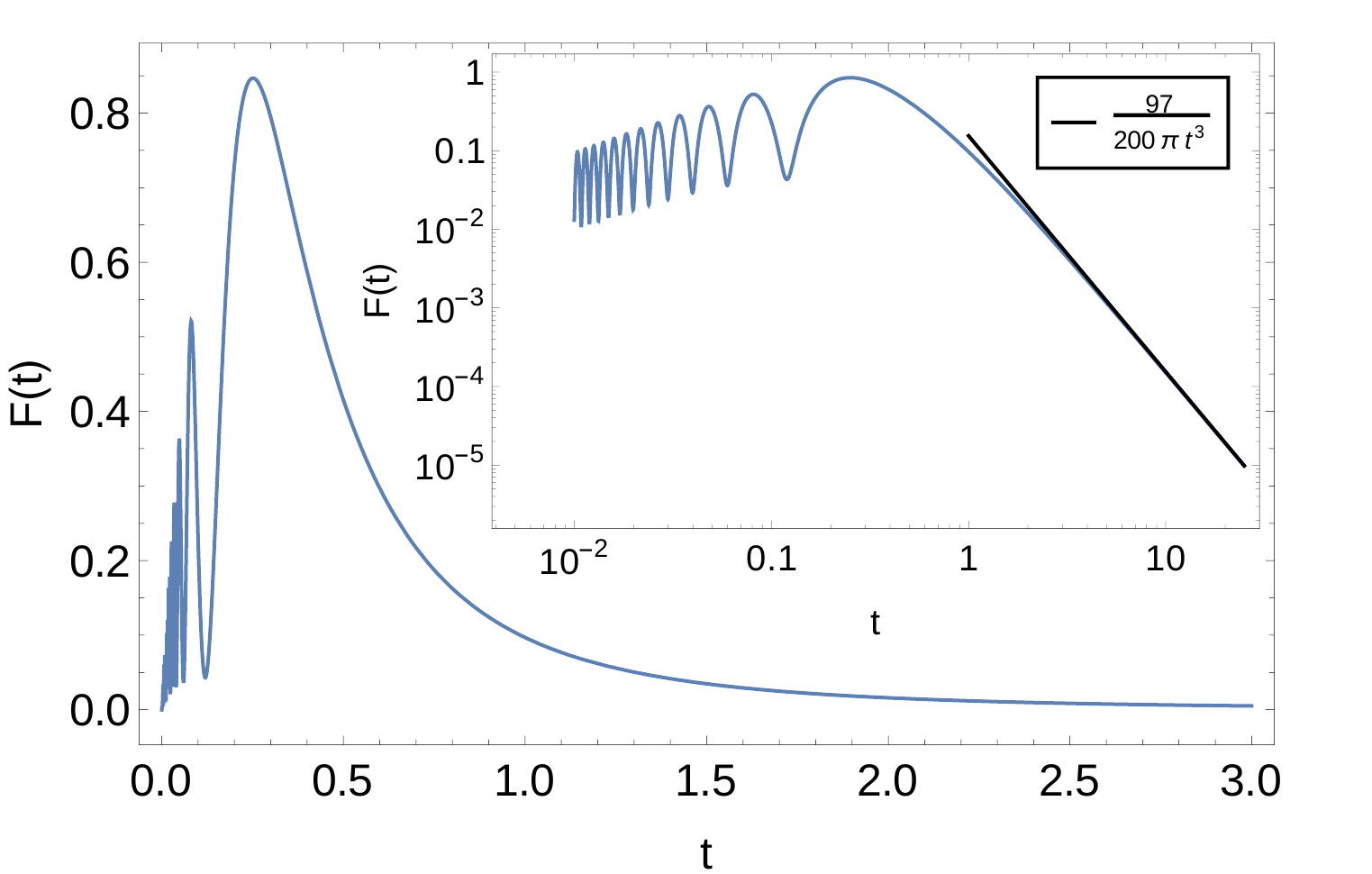}
	\caption{\label{fig:fptsq} The first passage time distribution $F(t)$ for the initial wavepacket and parameter values considered in Fig.~\eqref{fig:wavesq}. The inset shows the decay $F(t) \sim t^{-3}$ at large times, with a pre-factor given by Eq.~\eqref{eq:FT1}.}
\end{figure}

The survival probability $S(t)$ is given by  
\begin{equation}
S(t)=\int_{0}^{\infty}\left|\Psi(x , t)\right|^{2}\,dx\label{eq:Surv-Def}
\end{equation}
and after some straightforward manipulations one can show, using  Eq.~\eqref{eq:SE-cont} that
\begin{equation}
S(t)=1+2\frac{\Im(\zeta)}{\left|\zeta\right|^{2}}\int_{0}^{t}\left|\Psi(0 ,\tau)\right|^{2}\,d\tau
\end{equation}
which is the continuous counterpart of Eq.~\eqref{eq:SurvInteg}. For purely
real $\zeta$, i.e. $\alpha =0$ which corresponds to non-measurement, the system Eq.~\eqref{eq:SE-cont} determines unitary evolution
on half line. We note that the first passage time distribution is given by
\begin{equation}
\label{eq:fptd}
F(t)=- \cfrac{dS(t)}{d t} =-2\frac{\Im(\zeta)}{\left|\zeta\right|^{2}}\left|\Psi(0 , t)\right|^{2} \; >\; 0.
\end{equation}
From the asymptotic scaling form in Eq.~\eqref{eq:Asymptotic} we find that
\begin{equation}
\begin{gathered}
S_{\infty}=\lim_{t\to \infty} S(t) =\int_{0}^{\infty}\left|\frac{1-\imath\zeta k}{1+\imath\zeta k}\right|\left|c({k}) \right|^{2}dk.\label{eq:S-inf}
\end{gathered}
\end{equation}
We are not able to find more explicit forms for $F(t)$ or $S(t)$. However we can obtain the asymtotic long time form of $F(t)$. We need the wavefunction at the origin, $\Psi(0,t)$. This can be obtained from the scaling solution in Eq.~\eqref{eq:scal0} by use of the boundary condition $\Psi(0,t)= -\zeta \left[\tfrac{\partial\Psi}{\partial x} \right]_{x=0}$. This gives us, for $t \to \infty$, $\Psi(0,t)=
[m_{\Psi_0} {\zeta}/{\sqrt{4 \pi t^3}}] e^{-\imath {3\pi}/{4}}$. Hence we get
\begin{equation}
\label{eq:FT1}
\lim_{t\to \infty} t^3  F(t) =  -\f{\Im (\zeta)}{2 \pi} \vert m_{\Psi_0}\vert^2.
\end{equation}
In Fig.~\eqref{fig:fptsq} we plot  $F(t)$ for the same parameters and initial wavefunction used in Fig.~\eqref{fig:wavesq}. At large times, we verify the above asymptotic form given in Eq.~\eqref{eq:FT1}.

 We observe that the righthand side of \eqref{eq:FT1} may vanish for some special initial condition $\Psi_0$. This means that the asymptotic decay of the first passage time distribution is not universal.  An exhaustive study can be performed showing that it is always possible to start with a special initial wave function $\Psi_0$ such that the asymptotic decay of the first passage time distribution will be of order $t^{-(2s+1)}$ for some integer $s\ge 1$. The detailed study is performed in App.~\eqref{subsec:DecayFPTD}.

%\comment{The following should go in the appendix}}
%
%
%\medskip 
%
%{\bf Numerical examples}: Let us consider a few representative initial conditions.
%
%(a) We choose a flat initial wave function given by
%\begin{align}
%\psi_0(x)=1,~~{\rm for}~1<x<2, 
%\end{align}
%and zero elsewhere. We then find
%\begin{align}
%c_k= \frac{2 \sin (k/2) [k \zeta \cos (3 k/2) - \sin (3 k/2)]}{k \pi (1 + k^2 \zeta^2)},~~ c_b=\sqrt{2 \zeta} e^{-2/\zeta} (-1+e^{1/\zeta}).
%\end{align}
%
%
%
%(b) Our second choice is the smooth wave-function given by
%\begin{equation}
%\psi_{0}(x)=\begin{cases}
%\frac{2}{\sqrt{3}}x^{2}\exp(-x) & x\ge0\\
%0 & x<0
%\end{cases}.
%\end{equation}
%This gives
%\[
%c_{k}=\frac{4 k\left[\zeta -3 + (1-3\zeta) k^{2}\right]}{\sqrt{3} \pi \left(1+k^{2}\right)^{3}(1+\zeta^{2}k^{2})},~~c_b=4\sqrt{\frac{2}{3}} \frac{\zeta^{5/2}}{(1+\zeta)^3}.
%\]
%
%

%\begin{figure}
%\includegraphics[scale=0.4]{wavesq.pdf}
%\caption{\label{fig:wavesq}
%Plot showing  $|\psi(x,t)|^2$ at times $t=0,0.1,0.5$, obtained from the solution in Eq.~\eqref{gensol} with the square initial condition $\psi(x,0)=1$ for $1<x<2$ and zero elsewhere. The parameter value $\zeta=0.2+0.5 i$ was taken.} 
%\end{figure}
%
%
%\begin{figure}
%\includegraphics[scale=0.4]{fptsq.pdf}
%\caption{\label{fig:fptsq} The first passage time distribution $F(t)$ for the initial wavepacket and parameter values considered in Fig.~\eqref{fig:wavesq}. The inset shows a decay $F(t) \sim t^{-3}$ at large times.}
%\end{figure}

%{\color{blue} Question: Can one find the large time asymptotic form of $\psi(x,t)$. Perhaps there is a limiting scaling form?} 
%

\subsection{Real-line case}

The Schrodinger Eq. \eqref{eq:shro_Z} can be rewritten in the form
\begin{equation}
\label{eq:eqZ-RBC}
\begin{split}
& \imath\cfrac{\partial \psi_{n}}{\partial t}=2\psi_n -\psi_{n+1}-\psi_{n-1}, \quad n \in {\mathbb Z}\\
&\psi_0 = \imath w (\psi_{-1} + \psi_1 ).
\end{split}
\end{equation}

Assuming that $\imath w = \tfrac{1}{2}+\tfrac{\epsilon}{\zeta}$ where $\zeta$ is a complex number and defining the continuous wave function $\Psi$ as 
$$\Psi(x , t) = \lim_{t\to 0} \epsilon^{-1/2} \psi_{[x/\epsilon]} \big(t \epsilon^{-2}\big),$$
we get in the limit $\epsilon\to 0$ the above equation reduces  to the Schrodinger equation with a complex Robin boundary condition at the origin   
\begin{equation}
\label{eq:SE-Z-cont}
\begin{split}
&\imath \tfrac{\partial\Psi}{\partial t} =- \tfrac{\partial^2\Psi}{\partial x^2}, \\
&2 \Psi\vert_{{x=0}}+{\zeta}\left[\tfrac{\partial\Psi}{\partial x}\vert_{x=0^+} - \tfrac{\partial\Psi}{\partial x}\vert_{x=0^-} \right]=0.
\end{split}
\end{equation}
The solution of Eq. \eqref{eq:SE-Z-cont} can be obtained in the following way. Let us define
\begin{equation*}
\Psi^{s} (x,t) =\cfrac{\Psi(x,t)+\Psi (-x, t)}{2}, \quad \Psi^{a} (x,t) =\cfrac{\Psi(x,t)-\Psi (-x, t)}{2},
\end{equation*}
the symmetric and antisymmetric part of the wave function $\Psi$. Both functions are uniquely determined by their restrictions to the half line $[0,\infty)$. Eq. \eqref{eq:SE-Z-cont} implies that on $[0,\infty)$, $\Psi^s$ is solution of the Schrodinger Eq. \eqref{eq:SE-cont} with a complex  Robin boundary condition at the origin and that on $[0, \infty)$, $\Psi^a$ is solution of the Schrodinger equation with the Dirichlet boundary condition $\Psi^a (0, t)=0$. To solve the latter, we observe that it is in fact sufficient to solve the free Schrodinger equation on the real line with the initial antisymmetric wave function $\Psi^a$ since this property will be preserved by the free propagator and in particular the solution will vanish on $0$ at any time $t>0$. It follows that the solution of Eq. \eqref{eq:SE-Z-cont} is given by
\begin{equation}
\begin{split}
\Psi (x, t)&=\int_{0}^{\infty}dk\,\big[{\hat c}({k}) {\hat \eta}^{k}(x)+a({k}) \sigma^{k}(x)\big]\,e^{-\imath k^{2}t}+{\hat c}_{b}{\hat \eta}^{b}(x)\,e^{\imath\frac{t}{\zeta^{2}}},\\
{\text{where}}\quad  {\hat c}({k})& =\int_{-\infty}^{\infty}dx\, \Psi_{0}(x){\hat \eta}^{k}(x), \quad a({k})=\int_{-\infty}^{\infty}dx\, \Psi_{0}(x)\sigma^{k}(x),\\
{\hat c}_{b}& =\int_{-\infty}^{\infty} dx\, \Psi_{0}(x){\hat \eta}^{b}(x),
\end{split}
\end{equation}
with the scattering states ${\hat \eta}^k, \sigma^k$  and the bound state ${\hat \eta}^b$ defined by Eq. \eqref{eq:eta_k_R} and Eq. \eqref{eq:eta_b_R}. The latter are nothing but the eigenfunctions of the Laplacian operator on the real line with complex Robin boundary condition at the origin as it appears in Eq. \eqref{eq:SE-Z-cont}, see Appendix \eqref{subsec:Specrealline}.\\

The survival probability $S (t)$ is given by  
\begin{equation}
S (t)=\int_{-\infty}^{\infty}\left|\Psi(x , t)\right|^{2}\,dx
\end{equation}
and after some straightforward manipulations one can show, using  Eq. \eqref{eq:SE-Z-cont} that
\begin{equation}
S (t)=1+4\frac{\Im(\zeta)}{\left|\zeta\right|^{2}}\int_{0}^{t}\left|\Psi(0 ,t')\right|^{2}\,dt'.
\end{equation}
Since $\sigma^k (0)=0$ and ${\hat \eta}^k$, ${\hat \eta}^b$ are even, we conclude that the first passage time distribution $F=-{\partial S}/{\partial t}$ is up to a multiplicative constant the same as the one for the half-line, but starting from the wave function $\Psi_0^s$ (instead of $\Psi$) restricted to the half-line. We observe in particular that if we start with an antisymmetric wave function $\Psi=\Psi^a$ then the particle is never detected.

\section{Conclusion}
\label{sec:conclusions}
Subjecting a quantum particle to repeated projective measurements at regular intervals of time ($\tau$) to ascertain its observation by a detector is one of the standard methods to study the quantum time of arrival problem. Since the work of Allcock~\cite{allcock1969time}, it is believed that an equivalent way to study the first detection    problem is to introduce an imaginary absorbing potential. In the present work we  discussed a precise limiting procedure of taking $\tau \to 0$ and at the same time allowing the coupling strength between system and detector to be large (as $~1/\tau^{1/2}$). We summarize here our main results:  
\begin{itemize}
\item Formulating the problem for general quantum systems with a discrete Hilbert space, we rigorously showed the equivalence between the repeated measurement protocol and the non-Hermitian description.
\item  For a quantum particle on a 1D lattice with a detector at one site we then solved the corresponding Schrodinger equation with a complex potential to obtain closed form analytic results for the survival probability and the distribution of first detection time of a particle starting from an arbitrary initial lattice site. Various asymptotic cases were discussed. 
\item We then studied the limit of  lattice spacing going to $0$ to obtain a formulation for the continuum case. For the semi-infinite lattice with a detector at one end we find that the effective description is in terms of free Schrodinger evolution with complex Robin boundary conditions at the detector site. Again in this case we provide  analytic results for several objects of interest. The long time asymptotic  form of the surviving wave-function was obtained. We find that while the detection time probability density generically decays as $1/t^3$, it is possible to construct special initial states for which the decay is faster.
 A similar dependence of decay exponent on initial states was observed in a lattice study~\cite{thiel2018spectral} and it will be interesting to relate these results.  
\end{itemize}

We note that this non-Hermitian description on the  ${\mathbb N}$-lattice remains unchanged even if we consider the detector to be localized on all  sites on the negative axis. Taking the continuum space limit then gives us the case considered by Allcock~\cite{allcock1969time}. However, we see that for with our model of system-detector interactions, the non-Hermitian description is in terms of one with complex Robin boundary conditions instead of one with an imaginary  potential~\cite{Tumulka2019}. An interesting direction for future work would be to consider other forms of the system-detector interaction that could lead to different forms of effective non-Hermitian Hamiltonians. 
An extension of our  results  to higher dimensions would be another interesting question to explore.

\begin{acknowledgements}
VD and AD  acknowledge support of the Department of Atomic Energy, Government of India, under project no.12-R$\&$D-TFR-5.10-1100. C.B. acknowledges support of MITI-CNRS under the project StrongQU. This work has been supported by the project RETENU ANR-20-CE40-0005-01 of the French National Research Agency (ANR) and the French-Indian UCA project ``Large deviations and scaling limits theory for non-equilibrium systems.'' 
\end{acknowledgements}

\bibliographystyle{unsrt}
\bibliography{references}

\appendix
\section{Spectrum Analysis}
\label{sec:app-spec-g}

\subsection{Spectrum on the half-line}
\label{subsec:app-spec-halfline}
The operator appearing in Eq. \eqref{eq:SE-cont} has a complete set of  basis functions comprising of a continuum of scattering states given by
\begin{align}
\label{eq:eta_k}
\eta^k(x)=\cfrac{\imath}{\sqrt{2\pi (1+\zeta^2 k^2)}}\left[(1-\imath k \zeta) e^{\imath kx}-(1+\imath k \zeta) e^{-\imath kx}\right],\quad  k >0,
\end{align}
and, for $\Re [\zeta] >0$, a bound state given by 
\begin{align}
\label{eq:eta_b}
\eta^b(x) = \sqrt{\f{2}{\zeta}} e^{-x/\zeta}.
\end{align}
Here $\sqrt z$ denotes the principal square root of $z\in \mathbb C\backslash {\mathbb R}_{-}$ using the nonpositive real axis as a branch cut. They satisfy the boundary condition at $x=0$ and the ``real orthonormality conditions'':
\begin{equation}
\label{eq:orthonormality}
\begin{split}
&\int_0^\infty dx ~\eta^k(x) \eta^{k'}(x) =  \delta (k-k'), \\
&\int_0^\infty dx ~\eta^k(x) \eta^{b}(x) = 0, \\
&\int_0^\infty dx ~[\eta^b (x)]^2 =1.
\end{split}
\end{equation}
This is easily proved using the identity $\int_0^\infty dx e^{i k x}=i P(1/k)+\pi \delta(k)$, where $P$ denotes the principal part. One also has 
the completeness relation:
\begin{align}
\label{eq:completeness}
\begin{split}
\int_0^\infty dk ~\eta^k(x) \eta^{k}(x')  &= \delta (x-x')~~{\rm for}~\Re[\zeta]<0, \\
\int_0^\infty dk ~\eta^k(x) \eta^{k}(x') + \eta^b(x) \eta^{b}(x') &= \delta (x-x')~~{\rm for}~\Re[\zeta]>0.
\end{split}
\end{align}
To prove this, we note that
\begin{align}
&\int_0^\infty dk ~\eta^k(x) \eta^{k}(x') \nn \\
&= -\f{1}{2 \pi}\int_0^\infty dk \frac{1}{{ (1+\zeta^2 k^2)}}\left[(1-\imath k \zeta) e^{\imath kx}-(1+\imath k \zeta) e^{-\imath kx}\right]~\left[(1-\imath k \zeta) e^{\imath kx'}-(1+\imath k \zeta) e^{-\imath kx'}\right] \nn \\
&=\delta(x-x')  -\f{1}{2 \pi}\int_0^\infty dk \left[ \frac{(1-\imath k \zeta)^2 e^{\imath k(x+x')}}{ (1+\zeta^2 k^2)} +  \frac{(1+\imath k \zeta)^2 e^{-\imath k(x+x')}}{{ (1+\zeta^2 k^2)}}\right] \nn \\
&=\delta(x-x')  -\f{1}{2 \pi}\int_{-\infty}^\infty dk  \frac{(1-\imath k \zeta) e^{\imath k(x+x')}}{ (1+i k \zeta)} \nn \\
&=\delta(x-x')  -\f{1}{\pi}\int_{-\infty}^\infty dk  \frac{ e^{\imath k(x+x')}}{ 1+i k \zeta}, 
\end{align}
where we used the fact that $\int_{-\infty}^\infty dk e^{i k (x+x')}=0$ since $x,x'>0$.Performing the integral, we then get the completeness relations in Eq.~\eqref{eq:completeness}.

\subsection{Spectrum on the real line}
\label{subsec:Specrealline}

The operator appearing in Eq. \eqref{eq:SE-Z-cont} has a complete set of  basis functions comprising of a continuum of scattering states given by
\begin{equation}
\label{eq:eta_k_R}
\begin{split}
{\hat \eta}^{k}(x)&=\frac{\imath}{\sqrt{4\pi\left(1+\zeta^{2}k^{2}\right)}}\bigg[(1-\imath\zeta k)e^{\imath k\left|x\right|}-(1+\imath\zeta k)e^{-\imath k\left|x\right|}\bigg], \quad k>0, \\
\sigma^{k}(x)&=\frac{1}{\sqrt{\pi}}\sin\left(k\,x\right),\quad k>0. 
\end{split}
\end{equation}
and, for $\Re [\zeta] >0$,  a bound state given by 
\begin{equation}
\label{eq:eta_b_R}
{\hat \eta}^{b}(x)=\frac{1}{\sqrt{\zeta}}\exp\left(-\frac{\left|x\right|}{\zeta}\right).
\end{equation}
Observe that, up to a multiplicative constant, ${\hat \eta}^k$ and ${\hat \eta}^b$ are respectively the symmetrization of ${\eta^k}$ and ${\eta^b}$. The eigenfunctions $\sigma^{k}$ are odd while ${\hat \eta}^k, {\hat \eta}^b$ are even. All the functions ${\hat \eta}^{b},\,{\hat \eta}^{k}$ and $\sigma^{k}$ satisfy the boundary condition in Eq. \eqref{eq:SE-Z-cont}. Actually $\sigma^{k}$ satisfy the boundary condition somewhat trivially since they are smooth and $\sigma^k (0)=0$. All these eigenfunctions also satisfy the ``real orthonormality conditions'':
\begin{equation}
\label{eq:orthonormality_R}
\begin{split}
&\int_{-\infty}^{\infty} dx\, [{\hat \eta}^{b}]^{2}=1, \\
&\int_{-\infty}^{\infty}dx\, {\hat \eta}^{k}{\hat \eta}^{k'}=\int_{-\infty}^{\infty}dx\, \sigma^{k}\sigma^{k'}=\delta(k-k'),\quad k,k'>0, \\ 
&\int_{-\infty}^{\infty} dx\,  \sigma^{k}{\hat\eta}^{k'} =\int_{-\infty}^{\infty} dx\, \sigma^{k}{\hat \eta}^{b}=\int_{-\infty}^{\infty} dx\, {\hat \eta}^{k}{\hat \eta}^{b}=0
\end{split}
\end{equation}
and the completeness relation:
\begin{align}
\label{eq:completeness_R}
&\int_0^\infty dk ~[{\hat \eta}^k(x) {\hat \eta}^{k}(x') + \sigma^k (x) \sigma^{k'} (x)]= \delta (x-x'),~~ {\rm for}~~\Re[\zeta]<0 \\
&\int_0^\infty dk ~[{\hat \eta}^k(x) {\hat \eta}^{k}(x') + \sigma^k (x) \sigma^{k'} (x)] + {\hat \eta}^b(x) {\hat \eta}^{b}(x') = \delta (x-x'),~~~ {\rm for}~ \Re[\zeta]>0.
\end{align}
The results of Eq.~\eqref{eq:orthonormality_R} are easily checked by using that $\int_{-\infty}^\infty dx e^{\imath r x} =2\pi \delta (r)$. The completeness relation Eq.~\eqref{eq:completeness_R} of these eigenfunctions is a consequence of the completeness relation Eq.~\eqref{eq:completeness} of the $\{\eta^k\; \; k>0\}, \eta^b$ when restricted to the half line. Indeed, if $\Psi(x) $ is a square integrable wave function, we can write it as $\Psi =\Psi^s +\Psi^a$ where $\Psi^s$ (resp. $\Psi^a$) is the symmetric (resp. antisymmetric) part of $\Psi$. We can decompose $\Psi^s$ on $[0,\infty)$ on the basis of the $\eta^k, \eta^b$ thanks to the completeness relation Eq.~\eqref{eq:completeness}. Since $\Psi^s$, ${\hat \eta}^k$, ${\hat \eta}^b$ are symmetric, the decomposition also holds on $(-\infty, \infty)$ with $\eta^k$ (resp. $\eta^b$) replaced by ${\hat \eta}^k$ (resp. ${\hat \eta}^b$), up to some multiplicative constants. On the other hand we can decompose the function $\Psi^a$ on the standard Fourier basis $\Psi^a (x) = (2\pi)^{-1/2} \int_{-\infty}^\infty dk [{\mathfrak F} \Psi^a] (k) e^{-\imath k x}$ and by the oddness of $\Psi^a$ it follows that $\Psi^a (x) = -\imath \sqrt{{2}} \int_{0}^\infty dk [{\mathfrak F} \Psi^a] (k) \sigma^k (x)$. This concludes the proof of the completeness.

\subsection{Spectrum on the ${\mathbb N}$-lattice}
\label{subsec:sa-H_N}

We want to study the spectrum of the non-hermitian Hamiltonian $H_{\mathbb N}$ defined by Eq. \eqref{eq:effh-N}. We start to look for eigenfunctions $(\psi_n)_{n\ge 1}$  associated to the complex eigenvalue $E$:
\begin{equation}
\label{eq:disShrN}
\begin{split}
(2-\imath w)\psi_{1}-\psi_{2} & =E \psi_1,\\
2\psi_n -\psi_{n-1}-\psi_{n+1} & = E \psi_n.~~n>1.
\end{split}
\end{equation}
This is equivalent to solving the second equation for $n \geq 1$ with the boundary condition
\begin{equation}
\label{eq:bcdisN}
\psi_0 + \xi (\psi_1-\psi_0) =0, \quad \xi= \cfrac{-\imath w}{1-\imath w}.  
\end{equation}

The most general scattering solutions are of the form $\psi_n=A e^{\imath k n}+B e^{- \imath k n}$. This satisfies the second equation of Eq. \eqref{eq:disShrN}  with $E_s (k)=2 (1-\cos (k))=4 \sin^2 (k/2)$. Then the boundary condition Eq. \eqref{eq:bcdisN} above gives $B = -A \tfrac{1-\xi +\xi e^{\imath k}}{1-\xi + \xi e^{-\imath k}}$.\\

This gives us all the scattering solutions with  and we define for each $k \in (0,\pi)$ the eigenfunction $\eta^k$ associated to the eigenvalue $2(1- \cos (k))$ by  
\begin{equation}
\label{eq:etank}
\eta_n^k = \cfrac{\imath}{\sqrt{2\pi (1-\xi +\xi e^{\imath k}) (1-\xi +\xi e^{-\imath k})}} \,  \left[ (1-\xi +\xi e^{-\imath k}) e^{\imath k n} -(1-\xi +\xi e^{\imath k}) e^{-\imath k n}\right]
\end{equation}
for any $n\ge 1$.

We can then look for bound state solutions of the form $e^{-k n}$ with $\Re(k)>0$. In this case we find $2(1- \cosh (k))$ and using the boundary condition it gives $1-\xi^{-1}=e^{-k}$, which has a solution with  ${\Re} (k)>0$ only for $\left\vert 1-\xi^{-1}\right\vert <1$ or equivalently $|w|>1$. In this case we define the bound state $\eta^b$ associated to the eigenvalue $E_b=1+\xi^{-1} $ by
\begin{equation}
\label{eq:etanb}
\eta^b_n= \dfrac{(1-\xi^{-1})^n}{\sqrt{1-(1-\xi^{-1})^2}}
\end{equation}
for any $n\ge 1$.\\

``Real orthogonality relations'' and completeness property similar to Eq. \eqref{eq:orthonormality} and Eq. \eqref{eq:completeness} may be establish as in the continuum case.

\subsection{Spectrum on the ${\mathbb Z}$-lattice}
\label{subsec:sa-H_Z}

The spectrum of the non-hermitian Hamiltonian $H_{\mathbb Z}$ defined by Eq. \eqref{eq:effh-Z} can be deduced from the spectrum of $H_{\mathbb N}$ derived in the previous section. We recall that the equations of motion are given by Eq. \eqref{eq:eqZ-RBC}. Let $(\psi_n)_{n\ne 0}$ be an eigenstate associated to the eigenvalue $E$. It satisfies
 \begin{equation}
\begin{split}
(2-\imath w)\psi_{1}-\psi_{2}-\imath w \psi_{-1} & =E \psi_1,\\
(2-\imath w)\psi_{-1}-\psi_{-2}-\imath w \psi_{1} & =E \psi_{-1},\\
2\psi_n -\psi_{n-1}-\psi_{n+1} & = E \psi_n, \quad |n|\ge 2.
\end{split}
\end{equation}
Introduce the symmetric part $\psi^s =\Big( \tfrac{\psi_n +\psi_{-n}}{2}\Big)_{n\ne 0}$ and the antisymmetric part $\psi^a=\Big( \tfrac{\psi_n -\psi_{-n}}{2}\Big)_{n\ne 0}$ of $\psi$. The functions  $\psi^s$ satisfy Eq. \eqref{eq:disShrN} for $n\ge 1$ with $w$ replaced by $2w$ and is therefore eigenfunctions of $H_{\mathbb N}$ associated to the eigenvalue $E$. On the other hand, the function $\psi^a$ satisfies for $n \ge 1$ the equations 
\begin{equation*}
\begin{split}
2 \psi^a_{1} -  \psi^a_{2} &= E  \psi^a_{1},\\
2 \psi^a_{n}- \psi^a_{n-1} -  \psi^a_{n+1} &= E  \psi^a_{n}, \quad n\ge 2,
\end{split}
\end{equation*}
which can be equivalently written as 
\begin{equation*}
2 \psi^a_{n}- \psi^a_{n-1} -  \psi^a_{n+1} = E  \psi^a_{n}, \quad n\ge 1,
\end{equation*}
with the boundary condition $\psi^a_0=0$. Since $\psi^a$ is antisymmetric, this is equivalent to say that $\psi^a$ is an antisymmetric eigenstate for the discrete Laplacian on $\mathbb Z$. Observe moreover that $\psi^s$ and $\psi^a$ are ``real orthogonal'' since $\sum_{n \ne 0} \psi^a_n \psi_n^s =0$ due to the symmetry properties of the two functions. \\

By introducing $\zeta=\tfrac{-2\imath w}{1-2\imath w}$ we deduce that the spectrum of $H_{\mathbb Z}$ is composed of:
\begin{itemize}
\item Symmetric scattering states ${\hat \eta}^k$ for each $k \in (0,\pi)$ associated to the eigenvalue $2(1- \cos (k))$ and defined for $|n|\ge 1$ by  
\begin{equation}
\label{eq:etankZ}
{\hat \eta}_n^k = \cfrac{\imath}{\sqrt{4\pi (1-\zeta +\zeta e^{\imath k}) (1-\zeta +\zeta e^{-\imath k})}} \,  \left[ (1-\zeta +\zeta e^{-\imath k}) e^{\imath k |n|} -(1-\zeta +\zeta e^{\imath k}) e^{-\imath k |n|}\right].
\end{equation}
\item Antisymmetric scattering states $\sigma^k$ for each $k \in (0,\pi)$ associated to the eigenvalue $2(1- \cos (k))$ and defined for $|n|\ge 1$ by  
\begin{equation}
\label{eq:sknZ}
\sigma^k_n = \cfrac{1}{\sqrt{\pi}} \sin (kn).
\end{equation}
\item A symmetric bound state ${\hat \eta}^b$ associated to the eigenvalue $E_b=1+\zeta^{-1}$ if $\left\vert 1-\zeta^{-1}\right\vert <1$ or equivalently $|2w|>1$. It is defined  by
\begin{equation}
\label{eq:etanbZ}
{\hat \eta}^b_n= \dfrac{(1-\zeta^{-1})^{|n|}}{\sqrt{2(1-(1-\zeta^{-1})^2)}}.
\end{equation}
\end{itemize}

These eigenstates satisfy ``real orthogonality relations'' by construction and a completeness property similar to Eq. \eqref{eq:orthonormality_R} and Eq. \eqref{eq:completeness_R} may be establish as in the continuum case.

\section{Technical computations to estimate the survival probability}
\label{sec:tecnic}

\subsection{$\mathbb{N}$ Lattice}
\label{subsec:N}

\subsubsection{Computation of the expression given in Eq. \eqref{eq:LapPsi1}}
\label{subsubsec:integral}

We compute here the quantity
\begin{equation}
Z(w, n_0)=\imath \dfrac{\displaystyle \dfrac{2}{\pi} \int_{0}^{\pi}dk\,\dfrac{\sin\left(k\right)\sin\left(n_{0}k\right)}{\imath s-2(1-\cos(k))}}{1+\imath\dfrac{2w}{\pi}\displaystyle\int_{0}^{\pi}dk\,\dfrac{\sin^{2}k}{\imath s-2(1-\cos(k))}}
\label{eq:LapPsi11}
\end{equation}
where $w=\alpha +\imath \beta$.  We will assume that $s \in \mathbb C$ satisfies $\Im (s) \ne -2$ and $s \notin \imath {\mathbb R}$. 

The integral above can be evaluated by means of contour integration. We first note that 
\[
\int_{0}^{\pi}dk\,\dfrac{\sin\left(k\right)\sin\left(n_{0}k\right)}{\imath s-2(1-\cos (k))}=\dfrac{1}{8}\left[-I(s,n_{0}+1)+I(s,n_{0}-1)+I(s,-n_{0}+1)-I(s,-n_{0}-1)\right]
\]
where
\[
I(s,n)=\int_{-\pi}^{\pi}dk\,\dfrac{\exp(\imath nk)}{\imath s-2(1-\cos (k))}=- \imath\int_{\mathcal C}dz\,\dfrac{z^{n}}{z^{2}-(2-\imath s )z+1}.
\]
The integration contour ${\mathcal C}$ in the second integral above is the unit
circle. It is easily seen that $I(s,n)=I(s,-n)$ so that we need only to compute $I(s,n)$ for $n\ge 0$. The poles of the integrand are 
\begin{equation*}
z_\pm (s)= \imath \theta_\pm \left( \dfrac{s}{2} +\imath \right) 
\end{equation*}
where 
\begin{equation}
\label{eq:thetapm}
\theta_{\pm} (z) =-z \pm \sqrt{z^2+1}.
\end{equation}
We need first to know the positions of theses poles with respect to ${\mathcal C}$. 

\medskip

These two functions are
well defined analytic functions in the subdomain $\Omega$, defined
as the complex plane $\mathbb{C}$ where the imaginary axis has been
removed apart from the open segment of the imaginary axis between
$-\imath$ and $\imath$. We claim that $\left|\theta_{+}(z)\right|<1$
and $\left|\theta_{-}(z)\right|>1$ for $z\in\Omega_{+}=\Omega\bigcap\{z\in\mathbb{C}:\Re(z)>0\}$
and $\left|\theta_{+}(z)\right|>1$ and $\left|\theta_{-}(z)\right|<1$ for
$z\in\Omega_{-}=\Omega\bigcap\{z\in\mathbb{C}:\Re(z)<0\}$. Since
the proof of the two claims are similar we only prove the first one.
Notice that $\theta_{\pm}$ are analytic on the connected set $\Omega_{+}$
and satisfy 
\[
\theta_{+}(z)\cdot \theta_{-}(z)=-1,\,\,\dfrac{\theta_{+}(z)+\theta_{-}(z)}{2}=-z.
\]
Consider $z\in\Omega_{+}$. If $\theta_{+}(z)$ or $\theta_{-}(z)$ was belonging
to $\mathcal{C}$ then both would be because of the first relation
above. It would then follow that $z=-\imath\sin\left[\arg(\theta_{+}(z))\right]$
which is excluded since $z\in\Omega_{+}$. A similar ad absurdio argument
shows that $\theta_{-}(z)$ and $\theta_{+}(z)$ do not belong to the imaginary
axis. Since $\theta_{\pm}$ are analytic on the connected set $\Omega_{+}$,
so is $\theta_{\pm}(\Omega_{+})$. The imaginary axis being excluded from
$\theta_{\pm}(\Omega_{+})$, the domains $\theta_{\pm}(\Omega_{+})$ are included
into $\{z\in\mathbb{C};\Re(z)>0\}$ or into $\{z\in\mathbb{C};\Re(z)<0\}$.
Since $\theta_{+}(1)=\sqrt{2}-1>0$ and $\theta_{-}(1)=-\sqrt{2}-1<0$ we get
$\theta_{+}(\Omega_{+})\subset\{z\in\mathbb{C};\Re(z)>0\}$ and $\theta_{-}(\Omega_{+})\subset\{z\in\mathbb{C};\Re(z)<0\}$.
Similarly, since $\theta_{\pm}(\Omega_{+})\bigcap\mathcal{C}=\emptyset$
the domains $\theta_{\pm}(\Omega_{+})$ are included into the interior
of the unit disc or into the exterior of the unit disc. The values
above of $\theta_{\pm}(1)$ imply that $\theta_{+}(\Omega_{+})$ is included
in the interior of unit disc and $\theta_{-}(\Omega_{+})$ is included
in the exterior of unit disc. Hence we have that that if $\Re(s)>0$,
the pole inside the unit disc is $z_{+}(s)$.

\medskip

It follows by the Residues Theorem that
\begin{equation}
I(s,n)=-\imath\pi\frac{[z_{+}(s)]^{n}}{\sqrt{\left(\frac{s}{2}+\imath\right)^{2}+1}},\quad \Re(s)>0.
\end{equation}
This gives
\begin{equation}
\int_{0}^{\pi}dk\,\dfrac{\sin\left(k\right)\sin\left(n_{0}k\right)}{\imath s-2(1-\cos k)} = 
-\dfrac{\pi}{2}[z_{+}(s)]^{n_0},\quad {\Re}(s)>0.
\label{eq:CountourInt1}
\end{equation}
From Eq. \eqref{eq:LapPsi11} and Eq. \eqref{eq:CountourInt1}
\begin{equation}
Z(w, n_0) = 
-\imath\dfrac{[z_{+}(s)]^{n_0}}{1-\imath w [z_{+}(s)]}, \quad {\Re}(s)>0.
\label{eq:LPsi}
\end{equation}

\subsubsection{Integrale approximation}
\label{subsec:Integrale approximation}

Here we estimate the integral 
\begin{equation*}
\begin{split}
\int_{0}^{1}du\, \frac{u^{2n_0-2}(1-u^{2})(1+\left|w\right|^{2}u^{2})}{(1+\left|w\right|^{2}u^{2})^{2}-(2{\Im}(w)\,u)^{2}}
&=\int_0^\infty dt \, e^{-(2n_0 -2) t} g(t)\\
& =\dfrac{1}{2(n_0-1)}\int_0^\infty dt\, e^{-t} g \left(\dfrac{t}{2(n_0-1)}\right)  
\end{split}
%\label{eq:S_N1}
\end{equation*}
where
\begin{equation*}
g(t)= e^{-t} \cfrac{(1-e^{-2t})(1+\left|w\right|^{2} e^{-2t})}{(1+\left|w\right|^{2} e^{-2t})^{2}-(2{\Im}(w)\,e^{-t})^{2}}.
\end{equation*}
For $t$ near $0$ we have
\begin{equation*}
g(t) \sim \dfrac{2 (1+| w|^2)}{(1+|w|^2)^2 -4 [{\Im} (w)]^2} \, t
\end{equation*}
and therefore for large $n_0$ we get
\begin{equation}
\int_{0}^{1}du\, \frac{u^{2n_0-2}(1-u^{2})(1+\left|w\right|^{2}u^{2})}{(1+\left|w\right|^{2}u^{2})^{2}-(2{\Im}(w)\,u)^{2}} \sim \cfrac{1}{n_0^2} \; \dfrac{ (1+| w|^2)}{2(1+|w|^2)^2 -8 [{\Im} (w)]^2}.
\end{equation}

\subsubsection{Expression of $\psi_1$ in terms of Bessel functions of first kind}
\label{subsec:psi_1-expression}

We obtain here the explicit formula Eq. \eqref{eq:psi_1Bessel} for $\psi_1$ whose Laplace transform can be expressed as
\begin{equation*}
[{\mathcal L} \psi_1] (s) =-\imath^{n_0+1} \chi \left(\dfrac{s}{2} +\imath \right)
\end{equation*}
where
\begin{equation*}
\chi (z) =\dfrac{[\theta_{+} (z)]^{n_0}}{1 +w \theta_{+} (z)}
\end{equation*}
with the function $\theta_\pm$ defined in Eq. \eqref{eq:thetapm}. We recall that this function is analytic in the subdomain $\Omega$, defined as the complex plane $\mathbb{C}$ where the imaginary axis has been
removed apart from the open segment of the imaginary axis between
$-\imath$ and $\imath$. Moreover we have seen that
\begin{equation*}
\begin{split}
{\Re}(z)>0 \quad \text{implies} \quad |\theta_+ (z)|<1, \quad |\theta_- (z)|>1.
\end{split}
\end{equation*}

Let us first identify $\chi (z)$ as the Laplace transform $[{\mathcal L} f] (z)$ of some explicit function $f$. It is known that if $z\in \mathbb C$ satisfies ${\Re}(z)>0$ then
\begin{equation*}
\int_0^\infty dt e^{-zt}  \dfrac{J_k (t)}{t} = \cfrac{(\theta_+ (z))^k}{k}.
\end{equation*}
Assuming that $z$ is such that $|w\theta_+(z)|<1$ we have then that
\begin{equation*}
\begin{split}
\chi (z) &= \sum_{k=0}^\infty (- w)^k [\theta_+ (z)]^{n_0 +k} =\sum_{k=0}^{\infty} (k+n_0) (- w)^{k} \int_0^{\infty} dt e^{-zt} \dfrac{J_{k+n_0} (t)}{t} \\ 
&= \int_0^{\infty} dt e^{-zt} f(t) dt =[\mathcal L f](z) 
\end{split}
\end{equation*}
with
\begin{equation*}
f(t)= \sum_{k=0}^{\infty} (k+n_0) (- w)^{k}\dfrac{J_{k+n_0} (t)}{t}. 
\end{equation*}
The interversion of the sum and the integral is justified since $J_k (2t) \sim (2\pi k)^{-1/2} (e t /k)^k$ for large $k$. It follows that for any time $t\ge 0$
\begin{equation*}
\psi_1 (t) = -2 \imath^{n_0+1} e^{-2\imath t} \sum_{k=0}^{\infty} (k+n_0) (- w)^{k}\dfrac{J_{k+n_0} (2t)}{2t}
\end{equation*}
since the functions above have their Laplace transform coinciding on $\{s \in \mathbb C\; ; \; \Re (s) >\sigma\}$ for some $\sigma>0$ sufficiently large.

\subsection{Estimate of the asymptotic forms of the wavefunction and  of the first passage time distribution on the half line}
\label{subsec:DecayFPTD}

{\bf Saddle point approximation for the wavefunction}: 
To seek the saddle point approximation of 
\begin{align}
\Psi(x,t)&=\int_{0}^{\infty}dk\,c(k)\,\eta^{k}(x)\,\exp(-\imath\,k^{2}t) \\
&=\frac{\imath}{\sqrt{2\pi}}\Bigg[\int_{0}^{\infty}dk\,c(k)\frac{(1-\imath\zeta k)e^{\imath kx-\imath k^{2}t}}{\sqrt{1+\zeta^{2}k^{2}}}-\int_{0}^{\infty}dk\,c(k)\frac{(1+\imath\zeta k)e^{-\imath kx-\imath k^{2}t}}{\sqrt{1+\zeta^{2}k^{2}}}\Bigg] \\
&=\frac{\imath}{\sqrt{2\pi}}\int_{-\infty}^{\infty}dk\,c(k)\frac{(1-\imath\zeta k)}{\sqrt{1+\zeta^{2}k^{2}}}e^{\imath t(k\frac{x}{t}-k^{2})},
\end{align}
where we used the fact that $c(-k)=-c(k)$. We now use the following result~\cite{Miller}: 
\textit{Let $g(k)$ be complex valued, $I(k)$ be real valued functions
of the real variable $k$. Let $t>0$ and define the integral }
\begin{equation}
F(t)=\int_{-\infty}^{\infty}dk\,g(k)\,e^{\imath t\,I(k)}.
\end{equation}
\textit{If $k_{0}$ is a stationary point of $I(k)$ such that $\big[\frac{dI(k)}{dk}\big]_{k=k_{0}}=I'(k_{0})=0$
and $\big[\frac{d^{2}I(k)}{dk^{2}}\big]_{k=k_{0}}=I''(k_{0})\ne0$
then for large $t$, the contribution to $F(t)$ from $k_{0}$ is}
\begin{equation}
\Bigg[\frac{2\pi}{t\left|I''(k_{0})\right|}\Bigg]^{1/2}g(k_{0})e^{\imath t\,I(k_{0})+\imath\frac{\text{sign}(I''(k_{0}))\pi}{4}}+o(t^{-1/2}).
\end{equation}

In the current case
\begin{equation}
g(k)=c(k)\frac{(1-\imath\zeta k)}{\sqrt{1+\zeta^{2}k^{2}}}\,\,\,\,\&\,\,\,\,I(k)=k\frac{x}{t}-k^{2}.
\end{equation}
We get a unique saddle point at $k_0=x/(2t)$ and this then gives
\begin{equation}
\left[\Psi(x,t)\right]_{t\rightarrow\infty}\asymp\frac{\imath}{\sqrt{2t}}c\left(\frac{x}{2t}\right)\sqrt{\frac{1-\imath\zeta\frac{x}{2t}}{1+\imath\zeta\frac{x}{2t}}}e^{\imath\left(\frac{x^{2}}{4t}-\frac{\pi}{4}\right)}.
\end{equation}

\medskip

\noindent{\bf First detection distribution}: 
Recall Eq. \eqref{gensol} giving the expression of the solution of the Schrodinger equation on the half-line with complex Robin boundary condition and initial condition $\Psi_0 (x) = \Psi (x,0)$. To simplify we assume that $\Psi_0$ has compact support and has a bounded derivative defined almost everywhere. In particular all the moments of $\Psi_0$ and $\tfrac{\partial \Psi_0}{\partial x}$ are well defined. 
%\begin{equation}
%\Psi(x;t)=\int_{0}^{\infty}c_{k}\eta^{k}(x)e^{-\imath k^{2}t}dk\,+c_{b}\eta^{b}(x)e^{\imath\frac{t}{\zeta^{2}}}
%\label{eq:CompleteSolution}
%\end{equation}
%
After a change of variables we can write $\Psi (0,t)$ as
\[
\Psi(0,t)=\sqrt{\frac{2}{\pi}}\frac{\zeta}{t}\int_{0}^{\infty}\frac{y}{\sqrt{1+\frac{\zeta^{2}}{t}y^{2}}}c \Big({\tfrac{y}{\sqrt{t}}}\Big)\exp\left(-\imath y^{2}\right)\,dy\; +\; \sqrt{\frac{2}{\zeta}}c_{b}\exp\left(\imath\frac{t}{\zeta^{2}}\right),
\]
where
\[
c({k})=-\sqrt{\frac{2}{\pi(1+\zeta^{2}k^{2})}}k\,H(k),
\]

\[
H(k)=\int_{0}^{\infty}\underbrace{\bigg[\Psi_{0}(x)+\zeta\frac{\partial\Psi_{0}}{\partial x} (x)\bigg]}_{\Phi(x)}\frac{\sin\left(k\,x\right)}{k}dx.
\]
Up to factors, $H(k)$ is the sine transform of $\Phi(x)$. If $H(k)$
were to vanish identically, then $\Phi(x)\equiv0$ and $\Psi_{0}(x)$
would correspond to the bound state. In this situation $\Psi(0,t)$
evolves solely by the second term and the first passage time distribution $F(t)$ defined by \eqref{eq:fptd} has exponential decay for $\zeta$ lying in the fourth quadrant.

Assume $H(k)$ is not identically
$0$. Substituting for $c({k})$ one has 
\[
\Psi(0,t)\approx-\frac{2}{\pi}\frac{\zeta}{t^{\frac{3}{2}}}\int_{0}^{\infty}\frac{y^{2}}{1+\frac{\zeta^{2}}{t}y^{2}}H\left(\frac{y}{\sqrt{t}}\right)\exp\left(-\imath y^{2}\right)\,dy
\]
where the exponentially decaying term is neglected. By expanding $\sin\left(k\,x\right)$,
and switching {\footnote{The switching is justified since $\Psi_0$ has compact support so that the moments of $\Phi$ grows at most geometrically.}} the summation and integration for $\Phi(x)$ one can obtain an absolutely convergent series expansion for $H\left(\frac{y}{\sqrt{t}}\right)$. We get that
\begin{equation}
\begin{split}
H\left(\frac{y}{\sqrt{t}}\right)&=\sum_{s=0}^{\infty}\frac{(-1)^{s}}{t^{s}}\frac{y^{2s}}{(2s+1)!}\int_{0}^{\infty}x^{2s+1}\Phi(x)\,dx\\
&= \sum_{s=0}^{\infty}\frac{(-1)^{s}}{t^{s}}\frac{y^{2s}}{(2s+1)!} \left[M_{2s+1}-(2s+1) \zeta M_{2s}\right]
\end{split}
\end{equation}
where $M_s=\int_{0}^{\infty}x^{s}\Psi_{0}(x)\,dx$ the $s$th moment  of $\Psi_0$. In the second line we perform an integration by parts. This allows us to finally write down the series
\begin{equation}
\label{eq:PsiApprox}
\begin{split}
\Psi(0,t)& \approx-\frac{2}{\pi}\frac{\zeta}{t^{\frac{3}{2}}}\bigg[\sum_{s=0}^{\infty}\frac{(-1)^{s}}{t^{s}}\frac{ \left[M_{2s+1}-(2s+1) \zeta M_{2s}\right]    I_{s}(\frac{\zeta}{\sqrt{t}})}{(2s+1)!}\bigg]\\
& \approx -\frac{2}{\pi}\frac{\zeta}{t^{\frac{3}{2}}}\left[\sum_{s=0}^{\infty}\frac{(-1)^{s}}{ (2s+1)!\,  t^{s}} { M_{2s+1}\left(1-\tfrac{\zeta}{\zeta_s}\right)    I_{s}\Big(\tfrac{\zeta}{\sqrt{t}}\Big)}\right]
\end{split}
\end{equation}
where
\begin{equation}
I_{s}(z)=\int_{0}^{\infty}\frac{y^{2s+2}}{1+z^{2}y^{2}}\exp(-\imath y^{2})\,dy
\end{equation}
and
\begin{equation}
\zeta_s = \cfrac{1}{2s+1}\cfrac{M_{2s+1}}{M_{2s}} 
\label{eq:Defzetas}
\end{equation}
with the convention that $\zeta_s=\infty$ if $M_{2s}=0$. Noting that 
\[
\lim_{z\rightarrow0}I_{0}(z)=-\frac{\sqrt{\pi}}{4}e^{\frac{\imath\pi}{4}}\quad  {\text{and}} \quad \lim_{z\rightarrow0}I_{1}(z)=-\frac{3\sqrt{\pi}}{8}e^{-\frac{\imath\pi}{4}}
\]
one has then for large $t$
\begin{equation}
\Psi(0,t) = \frac{1}{2\sqrt{\pi}}\frac{\zeta}{t^{\frac{3}{2}}}\bigg[\Big(1-\tfrac{\zeta}{\zeta_0}\Big)M_{1}e^{\frac{\imath\pi}{4}}\; -\; \Big(1-\tfrac{\zeta}{\zeta_1}\Big) \frac{M_{3}}{4t}e^{-\frac{\imath\pi}{4}}\bigg] \; +\; O(t^{-5/2}).
\label{eq:AsympPsitwoterms}
\end{equation}
This proves Eq. \eqref{eq:FT1}.\\

\medskip

Consider the state
\begin{equation}
\Psi_{0}(x)=\Theta(x-2)-\frac{\Theta(x-1)+\Theta(x-3)}{2}+\imath\frac{\Theta(x-1)-\Theta(x-3)}{2}\label{eq:ExampleState}
\end{equation}
where $\Theta(x)$ is the Heaviside step function. Then 
\[
M_{1}=\frac{1}{2}+2\imath,\,\,\,\zeta_{0}=2-\frac{\imath}{2}
\]
\[
M_{3}=\frac{25}{4}+10\imath,\,\,\,\zeta_{1}=\frac{67}{82}-\frac{17i}{164}.
\]

\begin{figure}
	\includegraphics[scale=.58]{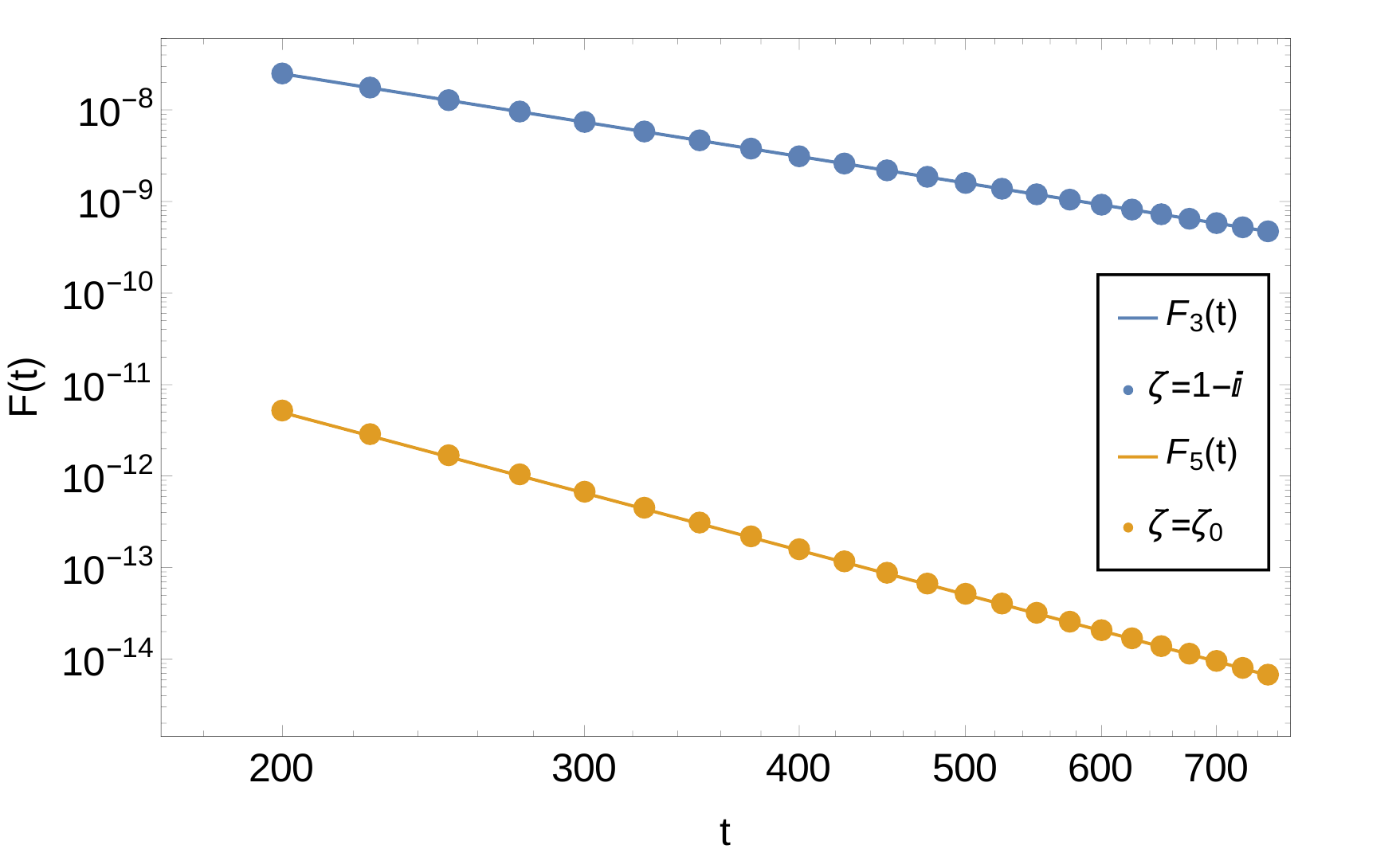}
	\caption{\label{fig:FNumvsAsymp} Plot comparing numerical estimates of $F(t)$ with $F_{3}(t)$ (resp. $F_{5}(t)$)   for $\zeta=1-\imath$ (resp. $\zeta(=\zeta_{0})=2-\imath/2$). The discrete points correspond to numerical values evulated from Eq.~\eqref{gensol}}.
\end{figure}

The numerical evaluation of the integral in Eq.~(\ref{gensol})
can be performed to obtain $\Psi(0,t)$. We choose the range $t\in(200,750)$ which is sufficient
to suppress the bound state contribution in Eq.~(\ref{gensol}).
To begin with let $\zeta=1-\imath$ . Then $\zeta$ is at a sufficient
distance from $\zeta_{0}$ and the first term in the expansion Eq.~(\ref{eq:AsympPsitwoterms})
dominates. Therefore one has as $t\to \infty$ that
\[
F_{3}(t)=-2\frac{\Im(\zeta)}{\left|\zeta\right|^{2}}\left|\Psi(0,t)\right|^{2}\sim \frac{5}{8\pi t^{3}}.
\]
We use this analytic estimate to compare with the numerical evaluation
via Eq.~(\ref{gensol}) in Fig $\eqref{fig:FNumvsAsymp}$. This shows that the analytic estimate is quite good and indeed $F(t)\sim\frac{1}{t^{3}}$. 

Now choose $\zeta=\zeta_{0}$. Doing so causes the first term in the
expansion Eq.~(\ref{eq:AsympPsitwoterms}) to drop out. The estimate for
$F(t)$ is obtained from Eq.~(\ref{eq:AsympPsitwoterms}) which gives
\[
F_{5}(t)\asymp-\frac{\Im(\zeta_{0})}{32\pi t^{5}}\left|M_{3}\Big(1-\frac{\zeta_{0}}{\zeta_{1}}\Big)\right|^{2}=\frac{5105}{1024}\frac{1}{\pi t^{5}}.
\]

Figure $\eqref{fig:FNumvsAsymp}$ shows good agreement of the above estimate with numerical values.

We could now claim that if $\Psi_{0}(x)$ was so constructed that
\[
\zeta_{0}=\zeta_{1}=\cdots=\zeta_{s-1}\ne\zeta_{s}(\ne0)
\]
then one has 
\[
F(t;\zeta=\zeta_{0})\sim\frac{1}{t^{3+2s}}
\]
while for other choices of $\zeta$ one has 
\[
F(t,\zeta\ne\zeta_{0})\sim\frac{1}{t^{3}}.
\]
This is under the assumption that the  moments $\int_{0}^{\infty}x^{k}\Psi_{0}(x)\,dx$
do not vanish up until $k=2s+1$. Lastly, we note that for the normalized state
\begin{equation*}
\Psi_{0}(x)=\sqrt{\frac{1}{\zeta_{0}}+\frac{1}{\zeta_{0}^*}}\exp\big(-\frac{x}{\zeta_{0}}\big)
\end{equation*}
with $\zeta_{0}$ in the fourth quadrant, $\zeta_{s}=\zeta_{0}$ for all $s$ follows from Eq. $\eqref{eq:Defzetas}$. For the measurement parameter $\zeta=\zeta_{0}$, the estimate in Eq. $\eqref{eq:PsiApprox}$ is identically $0$ and $F(t)$ falls exponentially, as has already been noted.

\end{document}